\documentclass[11pt,a4paper]{article}
\pdfoutput=1

\usepackage[T1]{fontenc}
\usepackage[utf8]{inputenc}
\usepackage[english]{babel}
\usepackage[a4paper,top=3cm,bottom=2.5cm,left=2.5cm,right=2.5cm,bindingoffset=0mm,
            marginparsep=7pt,marginparwidth=.6in,headheight=13.6pt]{geometry}

\usepackage{amsmath,amssymb,amsthm,mathtools}
\usepackage{appendix}
\usepackage{bm}
\usepackage[bitstream-charter]{mathdesign}
\DeclareSymbolFont{usualmathcal}{OMS}{cmsy}{m}{n}
\DeclareSymbolFontAlphabet{\mathcal}{usualmathcal}
\usepackage{mathrsfs,cancel,slashed}
\usepackage[subnum]{cases}

\usepackage{algorithm}
\usepackage{algpseudocode}

\usepackage{graphicx}
\graphicspath{{./plots/}}
\usepackage{xcolor}
\usepackage{float}
\usepackage{picinpar,xy}
\usepackage{tikz}
\usetikzlibrary{arrows.meta, positioning, fit, calc}
\usepackage{array}
\usepackage{booktabs}
\usepackage{multirow}
\usepackage{makecell}
\usepackage{adjustbox}
\usepackage{longtable}

\usepackage{xspace,ifthen,pifont,alltt,setspace,comment}
\usepackage{cite}
\usepackage[font=small,labelfont=bf,margin=0mm,labelsep=period,tableposition=bottom]{caption}
\usepackage[colorlinks=true, linkcolor=black!50!blue, urlcolor=blue, citecolor=blue, anchorcolor=blue]{hyperref}

\definecolor{red_cb}{HTML}{e41a1c}
\definecolor{blue_cb}{HTML}{377eb8}
\definecolor{green_cb}{HTML}{4daf4a}
\definecolor{purple_cb}{HTML}{984ea3}
\definecolor{orange_cb}{HTML}{ff7f00}

\definecolor{EmeraldGreen}{HTML}{1ea78d}
\definecolor{EnglishRed}{HTML}{b02427}
\hypersetup{
	pdftitle={Weight-Space-Programming-Omnilearn},
	pdfauthor={Bonnet-Guerrini et al.},
} 

\newcommand\one{\leavevmode\hbox{\small1\normalsize\kern-.33em1}}

\newcommand{\head}{\operatorname{head}}

\newcommand{\pT}{p_{\mathrm{T}}} 	
\newcommand{\pTjet}{p_{\mathrm{T},\,\mathrm{jet}}}

\newcommand{\arXiv}[2][]{%
	\ifthenelse{\equal{#1}{}}%
	{\href{http://arxiv.org/abs/#2}{arXiv:#2}}%
	{\href{http://arxiv.org/abs/#2}{arXiv:#2~[#1]}}}

\newcommand{\mev}{\text{MeV}}
\newcommand{\gev}{\text{GeV}}

\def\slashchar#1{\setbox0=\hbox{$#1$}           
   \dimen0=\wd0                                 
   \setbox1=\hbox{/} \dimen1=\wd1               
   \ifdim\dimen0>\dimen1                        
      \rlap{\hbox to \dimen0{\hfil/\hfil}}      
      #1                                        
   \else                                        
      \rlap{\hbox to \dimen1{\hfil$#1$\hfil}}   
      /                                         
   \fi}

\newcommand{\tikznode}[2]{%
\ifmmode%
\tikz[remember picture,baseline=(#1.base),inner sep=0pt] \node (#1) {$#2$};%
\else
\tikz[remember picture,baseline=(#1.base),inner sep=0pt] \node (#1) {#2};%
\fi}

\def\mathswitchr#1{\relax\ifmmode{\mathrm{#1}}\else$\mathrm{#1}$\xspace\fi}
\def\mathswitch#1{\relax\ifmmode#1\else$#1$\xspace\fi}

\begin{document}

\begin{center}{\Large \textbf{{
Weight-space programming and mechanistic interpretability \\for model editing in particle physics
}}}\end{center}

\vspace{0.3cm}
\begin{center}
Rapha{\"e}l Bonnet-Guerrini$^{1,2}$,
Johann Ioannou-Nikolaides$^{3}$, Troels Petersen$^{3}$
\end{center}

\begin{center}
{}$^1$Dipartimento di Informatica, Universit\`a di Milano,
Via Celoria 18, I-20133 Milano, Italy\\
{}$^2$INFN, Sezione di Milano, Via Celoria 16, I-20133 Milano, Italy and\\
{}$^3$Niels Bohr Institute, University of Copenhagen, Denmark

\vspace{0.9cm}

{\bf \large Abstract}
\end{center}
{
We demonstrate that weight-space programming and edits guided by sparse autoencoders can remove unwanted aspects of a particle-physics foundation model through edits to its weights. In a foundation model trained on jets, we observe that adding a single soft neutral particle causes large changes in jet classification, violating infrared safety. We locate the responsible attention heads with mechanistic analysis and trace the flaw's origin to pretraining on datasets with different reconstruction thresholds.
We then propose two targeted approaches to repair the flaw: by removing three latent directions identified with sparse autoencoders, and by letting language-model agents edit the weights directly, without gradients and with only 50 queries to the model. Both their edits individually remove almost the entire flaw and leave the rest of the model's response nearly unchanged. Compared to fine-tuning, both methods change far fewer parameters. Although built only on $W$ and QCD jets, the edits carry over to all signal classes, the pretrained generator, and a newly trained regression head.
}

\tableofcontents

\clearpage

\section{Introduction}

Analyses at the Large Hadron Collider~\cite{Evans:2008zzb,ATLAS:2008xda,CMS:2008xjf,LHCb:2008vvz,ALICE:2008ngc} must separate rare signals from large backgrounds in high-dimensional collision data. For a long time this was done with observables designed with physics intuition~\cite{Larkoski:2017jix}. Their definitions allow for a clear interpretation and calibration of algorithms and their uncertainties. The advent of deep learning allowed for direct training on the particles of jets and deep networks now outperform the physics observables~\cite{Kasieczka:2019dbj,Qu:2019gqs,Qu:2022mxj,Feickert:2021ajf}, but need to be retrained for each task on a dataset.

Foundation models~\cite{Bommasani2021OnTO} are a solution to this inefficiency. They are pretrained on large and diverse datasets, and can be adapted to many tasks with little labeled data and fine-tuning, often surpassing the performance of dedicated networks. Multiple foundation models have been proposed in jet physics~\cite{Birk:2024knn,Golling:2024abg,Mikuni:2024qsr,Mikuni:2025tar,Li:2024htp}, including OmniLearned~\cite{OmniLearned} which is pretrained on more than $10^9$ jets, and at the time of its introduction, set a new benchmark on anomaly detection tasks and jet tagging.

The shift from designed physical observables to large learned models brings substantial gains in predictive performance. However, it also makes their predictions less transparent. A physical observable has an explicit definition and a known response to changes in its inputs. A neural network builds its own internal representation that is shared across tasks and shaped by a pretraining objective that differs from the downstream analysis it is used for. Understanding which information it uses therefore becomes an important part of scientific validation.

Mechanistic interpretability~\cite{Olah2020ZoomIA,Elhage:2021framework,Bereska2024MechanisticIF, Sharkey2025OpenPI} seeks to identify the internal computations that produce a model’s behavior, the underlying representations that encode them, and tests their causality. Recent studies have applied these methods to high-energy physics, uncovering physically meaningful concepts in learned representations~\cite{Faucett2020MappingMP,Collins2021AnEO,Rai:2026uzm,Agarwal2026WhatDL,tiourine2026activation,UrquiaCalderon2026ZBoson,Wang:2024rup,Kitouni:2024ulw}. Because this approach locates where a semantic concept is computed, it also offers a route to correcting the mechanisms responsible for unwanted sensitivities.

Another major constraint of deep learning is domain shift, specifically in particle physics
when the detector response and reconstruction differ between simulation and data, and the datasets a foundation model is pretrained on~\cite{Barnard:2016qma,Nachman:2019dol,Ghosh:2021roe}. A model can then learn biases of the dataset rather than the physics~\cite{Geirhos:2020shortcut}.
Fine-tuning targets this failure mode~\cite{McCloskey1989CatastrophicII,Houlsby2019ParameterEfficientTL,Hu2021LoRALA}, but can lead to forgetting~\cite{McCloskey1989CatastrophicII,Kirkpatrick2016OvercomingCF} and does not locate the latents connected to it. Mechanistic analyses using e.g.\ sparse autoencoders (SAEs)~\cite{Bricken:2023monosemanticity,Huben2023SparseAF} identify interpretable features and intervene on them directly. Recently, weight-space programming (WSP) using reasoning language-model agents has been proposed to natively read model weights and fix unwanted flaws through direct interventions on model weights~\cite{takehi2026weightspace,Hu2026ExploringWP}.

Foundation models use shared representations across different downstream tasks, meaning that edits made to improve these representations can be inherited by all tasks.
\\
In this work, we investigate how the weights of a pretrained model can be edited directly, using OmniLearned as a case study.
\\
We first show that pretraining on multiple datasets with different reconstruction thresholds causes OmniLearned to violate infrared (IR) safety, altering jet encodings when a single soft neutral particle is added. We use mechanistic interpretability to diagnose this infrared sensitivity and localize its source within OmniLearned's shared representation.

We then compare three approaches to editing the model's weights to repair this sensitivity: low-rank fine-tuning, removing sparse-autoencoder latents, and weight-space programming using language-model agents. We find that all three approaches remove most of the infrared sensitivity for the specific tagger they are designed to repair, with improvements transferring across all downstream tasks tested.

Finally, we demonstrate that language-model agents can directly inspect and edit the weights of physics foundation models without gradients, using at most 50 queries, across three model sizes containing $3M$, $60M$, and $0.4B$ parameters. We further find that agent performance improves when the agents are provided with information about the targeted model behavior.

\section{Dataset Artifacts in Foundation Models}

Foundation models in particle physics are often pretrained on mixtures of datasets. These datasets  differ in simulation, detector responses and reconstruction, and foundation models will inevitably learn the distinct signatures of each~\cite{Geirhos:2020shortcut}. When a model relies on these non-physical artifacts to build its internal representation, it risks violating fundamental physics constraints in downstream tasks.
To investigate how these domain shifts manifest and how they can be mechanically repaired, we examine the class-specific jet taggers defined from OmniLearned's pretrained classification head~\cite{OmniLearned}, and the $W$-boson explicitly. We demonstrate that differences in pretraining datasets cause the OmniLearned backbone to learn a non-physical shortcut, directly violating infrared safety.

\subsection{An infrared sensitivity in OmniLearned's taggers}
\label{sec:wtag}
OmniLearned~\cite{OmniLearned} is a transformer-based model pretrained on jets from seven datasets. It comes in three different sizes ($-s$, $-m$, $-l$). For simplicity the exact numbers discussed in the paper correspond to its smallest version; we have confirmed that the infrared flaw and its origin are the same in all three versions, although in the larger models the response is spread over more attention heads (Sec.~\ref{sec:SAE_edits}).
The smallest version of OmniLearned has $3.3M$ parameters. Each particle is embedded into a 128-dimensional vector and a local embedding adds information about its nearest neighbors. Then, four learned summary tokens and an additional time-embedding token for the generative task are prepended to the particles. The information is passed through eight transformer blocks, with eight attention heads of dimension 16. In every block, the attention between two particles receives an additive bias that is computed by a network from the logarithms of the invariant masses, angular distance $\Delta R$ and relative transverse momentum $k_T$, as in~\cite{Qu:2022mxj}. It is pretrained with two heads, a generator head, and a classification head that predicts 200 specific jet flavor classes and 10 classes connected to the specific pretraining datasets used~(see Sec.~III in~\cite{OmniLearned} for a discussion of the datasets used). Without any further fine-tuning, we use the outputs for the JetClass~\cite{Qu:2022mxj} classes $W\to q\bar q$ and QCD to define a $W$ tagger,
\begin{equation}
  F_W = \ln\frac{p(W)}{p(\mathrm{QCD})},
\end{equation}
and select jets with $F_W$ above a threshold that retains 80\% of $W$ jets. On held-out JetClass jets with
$\pTjet$ between 500 and 1000~\gev, this selection accepts 2.0\% of QCD jets.

\paragraph{Infrared Safety}
\label{sec:irc}

\begin{figure}[t]
  \centering
  \includegraphics[width=0.7\textwidth]{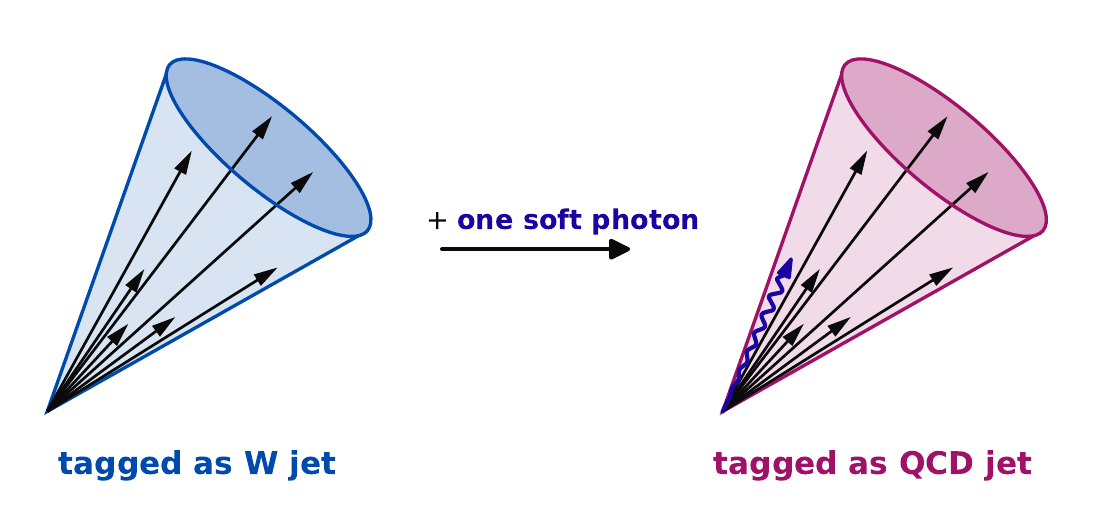}
  \caption{An infrared-safe tagger does not change its prediction when a soft particle is added. We observe for all pretrained OmniLearned taggers ($W$/$Z$/$H$/$t$ vs QCD) that adding a soft photon carrying $0.1\%$ of jet $\pT$ leads to the tagger rejecting the $W$ jet hypothesis for about a third of the $W$ jets it previously selected (Sec.~\ref{sec:wtag}).}
  \label{fig:sketch}
\end{figure}

Within any jet, the constituent particles carry varying fractions of the total momentum, with the softest falling near or below detector reconstruction thresholds. A physically meaningful jet observable must be robust against these soft emissions. The property that the change of its output vanishes as the energy of an added particle goes to zero is known as infrared (IR) safety~\cite{Sterman:1977wj}.
If the observable is also robust against collinear splittings, it is calculable in perturbative QCD. Jet clustering algorithms such as anti-$k_t$~\cite{Cacciari:2008gp} and observables used in jet tagging, such as the jet mass or N-subjettiness~\cite{Thaler:2010tr}, satisfy the condition by construction. A tagger built on them, such as a simple Boosted Decision Tree, gives the same answer for a jet with and without an additional soft particle.

For a learned neural network, this property is not guaranteed by construction~\cite{Komiske:2018cqr}. A network learns its response to soft particles from those in its training data, which can differ the most across datasets depending on the details of the detector simulation and the reconstruction thresholds. If they are not the same in all datasets a foundation model is pretrained on, a particle below the threshold of one dataset may then only appear in one part of the training data and the network may attend to the soft particles for dataset provenance classification.

Infrared safety therefore has to be tested. A finite set of tests cannot establish it, since it concerns the limit of vanishing energy, but it can show that a tagger violates it. We add a single particle with a fraction $f$ of the jet's $p_T$ at a random position within $\Delta R<0.4$ around the jet axis. For an infrared-safe tagger, the change of its output has to fall toward zero as $f\rightarrow0$. We find that the $W$ tagger of the pretrained OmniLearned does not satisfy this condition\footnote{It also responds to collinear splittings, but more weakly, and through a different mechanism which we leave for future work.}.

\begin{figure}[t]
  \centering
  \begin{minipage}{0.5\textwidth}\centering
    \includegraphics[width=\textwidth]{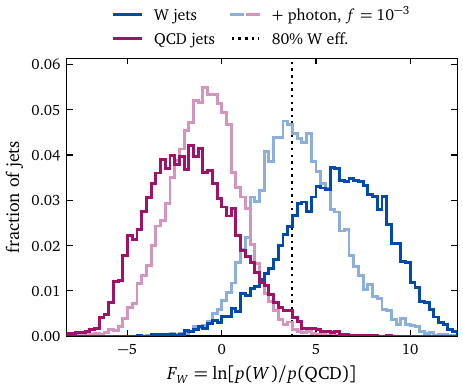}
  \end{minipage}\hfill
  \begin{minipage}{0.5\textwidth}\centering
    \includegraphics[width=\textwidth]{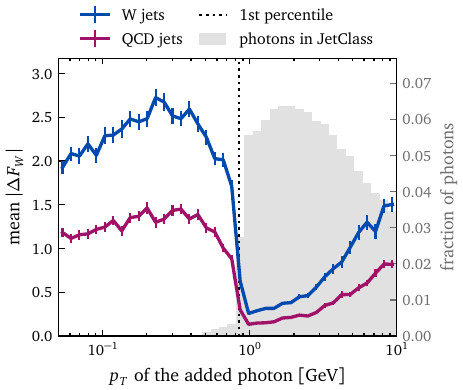}
  \end{minipage}
  \caption{Left: distribution of the W-tagger score $F_W$ for $W$ and QCD jets (dark) and for the same jets with one added photon at $f=10^{-3}$ (light). The vertical line indicates the threshold that retains 80\% of $W$ jets for the clean jets without an added photon. Right: mean change of $F_W$ for $W$ and QCD jets when one photon is added, as a function of its $\pT$. The gray histogram is the $\pT$ distribution of the photons in JetClass jets; the dotted line marks the first percentile.}
  \label{fig:hook}
\end{figure}

\paragraph{IR sensitivity of OmniLearned}
Adding one photon with $f=10^{-3}$ lowers the $W$ efficiency of this selection from $80\%$ to $54\%$ and the QCD acceptance from $2.0\%$ to $0.7\%$ (Fig.~\ref{fig:hook}, left). For a neutral hadron at $f=10^{-3}$ the $W$ efficiency falls to $46\%$, and the QCD acceptance to $0.6\%$. In the following, we state the impact of adding a soft photon in the main text, and report the numbers and figures for a soft neutral hadron in Appendix~\ref{app:figs}. This has a noticeable effect on a measurement of the $W$ fraction of a pseudo-dataset. We fit a pseudo-dataset of $2500$ $W$ and $7500$ QCD jets with two templates of $F_W$ obtained from simulation on the nominal pseudo-data. While the true $W$ fraction of a quarter is recovered for the nominal data, the fit on the tampered data has its best fit value at only $12\%$ with $\chi^2/\mathrm{ndf} = 1892/39$. A measurement made with this tagger therefore depends on soft particles with $f=10^{-3}$.

Fig.~\ref{fig:hook} (right) shows that the response of the $W$ tagger to soft photons is not a smooth function of the photon's momentum. From 10~\gev\ to 0.85~\gev\ it falls slowly, as expected. Below 0.85~\gev\ the response jumps up significantly, which coincides with the softest photon in JetClass for $|\eta|<1.5$, and remains high for photons outside the training data distribution. Neutral hadrons have a similar step in their response at their own threshold, whereas charged hadrons, electrons and muons do not (Appendix~\ref{app:figs}, Fig.~\ref{fig:app_types}), which we explain next.

\subsection{Locating the infrared sensitivity mechanistically}
\label{sec:IR_location}
\begin{figure}
  \centering
  \begin{minipage}{0.47\textwidth}\centering
    \includegraphics[width=\textwidth]{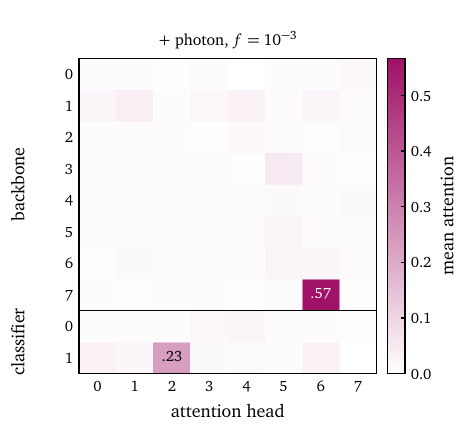}
  \end{minipage}\hfill
  \begin{minipage}{0.47\textwidth}\centering
    \includegraphics[width=\textwidth]{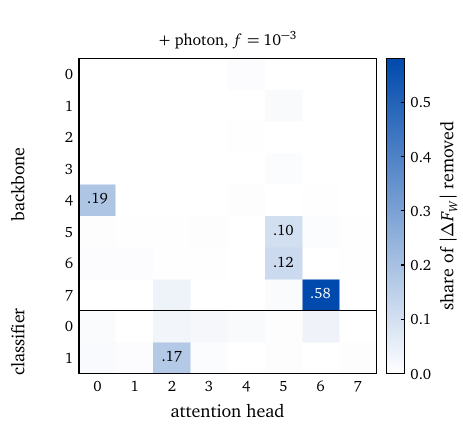}
  \end{minipage}
  \caption{Left: mean attention of the summary tokens on one added photon at $f=10^{-3}$ for every head of the backbone. Right: share of $|\Delta F_W|$ removed when the output of one head is restored to its value on the clean jet,
  for one added photon at $f=10^{-3}$. \texttt{b7/6} and \texttt{c1/2} attend strongly to the particle through their summary tokens, while \texttt{b4/0} through the particle tokens.}
  \label{fig:patching}
\end{figure}

In every attention head, a softmax over all tokens of the jet normalizes each summary token's attention weights so that they sum to one. Each attention head assigns a fractional weight to every particle in the jet, normalized to sum to one. The weight on the added particle therefore measures exactly how much of the head's output is driven by it.
Comparing this share across heads (Fig.~\ref{fig:patching}, left), we find that head 6 of transformer block 7 (abbreviated as \texttt{b7/6}), gives an added photon at $f=10^{-3}$ on average $57\%$, and $72\%$ at $\pT=0.3~\gev$ of the summary-token attention. No other head of the backbone gives it more than 5\%, and from the mean of particles in the jet dataset, the expectation would be no more than $2\%$ if attention were $\pT$-independent.

While the attention map shows where the particle is read, it does not show where this information is used. To test this, we use activation patching~\cite{Meng:2022rome,Zhang2023TowardsBP}.
We replace the output of each head for a jet with an added particle with the output of the clean jet, and measure how much of the change $|\Delta F_W|$ disappears.
We find that for a photon at $f=10^{-3}$, replacing the output of \texttt{b7/6} with the output of the clean jet removes $58\%$ of the $|\Delta F_W|$ response, while two further heads, head 0 of transformer block 4 (\texttt{b4/0}) and head 2 of the second transformer block of the classifier head (\texttt{c1/2}) carry most of the remaining response at $19\%$ and $17\%$ respectively. No other head carries more than $12\%$ (Fig.~\ref{fig:patching}) and their response depends on \texttt{b7/6}, \texttt{b4/0} and \texttt{c1/2}. We conclude that OmniLearned \emph{reads} soft particles with three heads specifically and similar attention patterns.

\paragraph{Characterizing the sensitivity}
We find that the readers \texttt{b7/6}, \texttt{b4/0}, and \texttt{c1/2} respond to two properties of the added particles. First, the readers only respond to neutral particles. In the JetClass format, charged particles carry the impact parameters of their tracks and their uncertainties, while all four entries are zero for neutral particles. Replacing the impact parameters of a soft photon by the uncertainties of a real track reduces the mean $|\Delta F_W|$ of $W$ jets from $2.4$ to $0.7$ and halves the attention of \texttt{b7/6} to these photons. Similarly, a soft charged particle with zero impact parameter uncertainty gets more attention.
Second, the response has a particle-type-dependent threshold. For photons, it lies at $\pT\approx 0.85$ GeV, around the same value as the softest photons in JetClass, while for neutral hadrons, we find it to lie at an energy about 1.9~\gev, just below the softest neutral hadron in JetClass at $E=2.3~\gev$ (Appendix~\ref{app:figs}, Fig.~\ref{fig:attention}, bottom).

\paragraph{A hypothesis for the origin of the sensitivity}
\label{sec:hypothesis}
We hypothesize that the pretraining objective of OmniLearned makes its JetClass $W$ tagger react to calorimeter deposits that do not occur in the JetClass dataset, but do occur in JetClass-II~\cite{Li:2024htp}, a second dataset used in pretraining. JetClass-II encodes neutral particles in the same way as JetClass and is produced with the same simulation framework but lower reconstruction thresholds ($71\%$ of JetClass-II jets have at least one neutral particle below JetClass's thresholds).
OmniLearned always ($99\%$ probability) correctly identifies a clean $W$ jet from the JetClass dataset, but after adding a photon at $0.3~\gev$ (below the JetClass calorimeter threshold) this probability drops to $1.5\%$, while the JetClass-II probability rises to $98.5\%$ (we find a similar response for soft neutral hadrons). One added particle is thus enough to change the dataset the model assigns the jet to.

To locate where in the network the dataset reassignment happens, we train a logistic-regression probe~\cite{Alain:2016probe} on the summary tokens after each layer, including the two transformer blocks of the classifier head, to estimate the probability that a jet comes from JetClass-II. For a JetClass jet with one added photon at $0.3~\gev$, the JetClass-II probability fluctuates around $0.3$ up to the sixth transformer block of the shared backbone, and follows that of a true JetClass-II jet after block 7 and for all blocks of the classifier head (see Appendix~\ref{app:figs}, Fig.~\ref{fig:probes}). As a control we check that adding a soft neutral particle to a JetClass-II jet does not change the probabilities.

\begin{figure}[t]
  \centering
  \includegraphics[width=\textwidth]{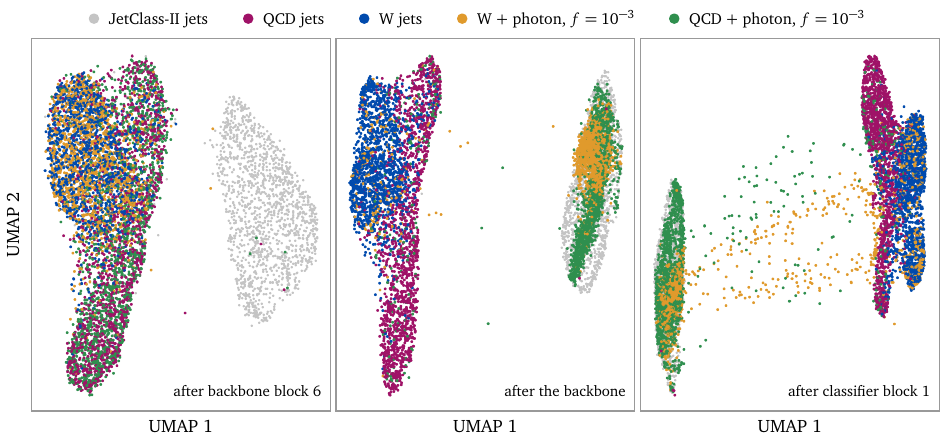}
  \caption{UMAP projection of the summary tokens of OmniLearned-s after transformer block 6 (left), after the shared backbone (middle), and after the last classifier block (right). The projection is fitted on clean JetClass $W$ and QCD jets and clean JetClass-II jets, and applied to the same $W$ and QCD jets with one added photon carrying a fraction $f=10^{-3}$ of the jet $\pT$.}
  \label{fig:umap}
\end{figure}

The same reassignment is visible in the latent space (Fig.~\ref{fig:umap}). We project the summary tokens with UMAP~\cite{McInnes:2018umap}, fitted on clean JetClass $W$ and QCD jets and clean JetClass-II jets only, and then apply the projection to the same $W$ and QCD jets with one added photon. After transformer block 6, the closest jets to the jets with the added photon are the clean JetClass jets of their class. After the shared backbone of OmniLearned, these are predominantly JetClass-II jets.
We also tested OmniLearn~\cite{Mikuni:2025tar}, which was only trained on the JetClass dataset~\cite{Qu:2022mxj}, and found no infrared sensitivity to soft neutral particles.

These observations confirm that during multi-dataset pretraining, the model consistently learns that a soft, neutral particle below JetClass's reconstruction threshold is a distinct signature of a JetClass-II jet. In the latent space, the presence of this single soft particle outweighs all other evidence and the jet is internally relabeled as JetClass-II, leaving the JetClass $W$ and QCD outputs unconstrained. The infrared sensitivity of the tagger is therefore a dataset-provenance shortcut~\cite{Geirhos:2020shortcut}, not a property of jet physics. Crucially, because this shortcut hijacks the model's shared internal representation, as the middle panel of Fig.~\ref{fig:umap} indicates, the flaw is not isolated to the $W$ tagger. We verify that all nine OmniLearned taggers besides the QCD tagger respond in the same way to one soft photon at $f=10^{-3}$ (Sec.~\ref{sec:taggers}), with signal efficiencies falling to between $12\%$ and $61\%$.
\\
The pretrained generator head is affected too: adding one photon of $0.3~\gev$ to a lightly noised $W$ jet changes the mass of the generated jet by $1.1\%$, almost three times the effect of a photon of $3~\gev$, above the threshold ($0.4\%$), and more than half the effect of changing the class label of the jet to QCD at the same step ($1.8\%$, Sec.~\ref{sec:heads}).
\\
Additionally, a new linear, downstream, jet mass regression head, trained on the frozen representation of the backbone, is also affected by the soft neutral particle sensitivity. Its prediction changes by $10.6~\gev$ averaged over 3000 jets from each of the ten classes, about twice its resolution, when a photon with $\pT$ equal to $0.1\%$ of the sum of the constituent $\pT$s is added, which changes the true jet mass by only $0.2~\gev$. The photon mostly pulls the light, two-pronged jets down in predicted mass ($-19.8~\gev$ for QCD jets and $-13.1~\gev$ for $W$ jets).

This illustrates a fundamental challenge for foundation models in physics: broad pretraining can instill dataset artifacts in the internal representation that can be causal features for downstream tasks. Fixing this necessitates targeted edit mechanisms to correct the representation without destroying the model's useful physics knowledge.

\section{Editing OmniLearned}
\label{sec:approaches}

In this section, we test whether we can reduce the sensitivity to soft, neutral particles identified in Sec.~\ref{sec:irc} while preserving the model's useful responses.
We propose two targeted approaches: a \textbf{mechanistic interpretability} driven approach that performs targeted interventions guided by SAEs; \textbf{weight-space programming} by language-model agents, which we give different levels of information. We compare both with \textbf{fine-tuning} with a dedicated objective to minimize the flaw, which we describe first.
For each approach, we compare different variants and then compare the selected variants across approaches, using the same metrics in all tables.
\paragraph{The fine-tuning baseline}
\label{sec:fine-tuning}

A common approach to adapting foundation models to specific tasks is fine-tuning~\cite{Bommasani2021OnTO}, which consists of updating the pretrained weights by minimizing a new objective. This optimization can be restricted to parts of the model to reduce the number of parameters changed and preserve the original behavior of the model~\cite{Houlsby2019ParameterEfficientTL,Hu2021LoRALA}. For our fine-tuning repair, we introduce a loss function aimed at removing the sensitivity of the $W$ tagger to soft neutral particles, while keeping the original predictions of the model intact
\begin{equation}
    \mathcal{L}_{\mathrm{FT}}
    = \mathcal{L}_{\mathrm{hold}}
    + \lambda \mathcal{L}_{\mathrm{IC}}
    + \mu \mathcal{L}_{\mathrm{phys}}.
    \label{eq:ft_loss}
\end{equation}
Here $\mathcal{L}_{\mathrm{hold}}$ penalizes changes in the predictions on unmodified jets, $\mathcal{L}_{\mathrm{IC}}$ penalizes the sensitivity to soft-particle additions, and $\mathcal{L}_{\mathrm{phys}}$ preserves the model's response to other physical changes in the jet. The latter include changes in the angular spread, particle momenta, and fragmentation pattern. For the last two terms, we use centered JetClass logits and normalize their differences by the typical variation between independent clean jets. The full definitions are given in Appendix~\ref{app:fine-tuning}.

In Sec.~\ref{sec:IR_location} we saw that jets with an added soft neutral particle are reassigned from JetClass to JetClass-II in the body of the model. To repair this shared representation, we update only the body's attention modules and the network computing the pairwise attention bias, using low-rank adaptations~\cite{Hu2021LoRALA}; all other weights stay frozen. We train on both photons and neutral hadrons, since training on photons alone leaves a larger response to neutral hadrons. Updating the whole body and classifier instead also removes most of the response, but more than halves the response to physical changes (Appendix~\ref{app:fine-tuning}). This is an example of catastrophic forgetting~\cite{McCloskey1989CatastrophicII, Kirkpatrick2016OvercomingCF}: gradient descent can change any behavior the loss does not protect. The selected setup passes all our checks, but fine-tuning gives no guarantee for behaviors we did not test. The same recipe also repairs the medium and large models, which are affected even more than the small one (one added soft photon drops their $W$ efficiency to $28\%$ and $40\%$).

\subsection{The sparse autoencoder approach}
\label{sec:SAE_edits}
SAEs are used in mechanistic interpretability to isolate human-understandable concepts into single neurons~\cite{Huben2023SparseAF}. This is not possible with the model's own neurons due to polysemanticity~\cite{Elhage:2022superposition,Bricken:2023monosemanticity}.
By encoding a layer's activations in a higher-dimensional sparse representation and reconstructing them, SAEs can spread concepts into single neurons.
\\
We train three TopK sparse autoencoders~\cite{Gao:2024topk} on the particle tokens at the input of each of the readers identified in Sec.~\ref{sec:IR_location}. To train them, we use $2.5M$ tokens from JetClass and JetClass-II jets, with $4\,096$ latents of which $32$ are active per token (TopK objective). The most responding latent, identified at \texttt{b7/6}, fires on all tested added sub-threshold neutral particles and on $99\%$ of the sub-threshold neutral particles in JetClass-II, but only on $0.8\%$ of real JetClass neutral particles. We find a similar latent in the other two attention heads \texttt{b4/0} and \texttt{c1/2}.
\\
Each latent corresponds to one direction in the input of its reader head. To edit the model, we remove this direction from the weights that the head uses to decide which particles to attend to. This is a rank-one change of one weight matrix per reader.
We sequentially remove all three readers (Appendix~\ref{app:mec-interp}). Editing all three readers results in the smallest remaining response of $0.11$ with $1\,533$ changed parameters and leaves the accuracy on unmodified jets unchanged. Its weak point is the response to physical changes, which falls to $0.78$. Editing fewer readers, or removing only half of each direction, keeps more of the physical response but repairs less (Appendix~\ref{app:mec-interp}, Table~\ref{tab:sae_variants}).\\
Following recent work by Cammarata~\cite{cammarata2026clean}, we also test whether single neurons of the model, either a hidden unit of an MLP or one of the components of a particle's vector, isolate the sub-threshold neutral particles without an SAE. We find that some single neurons carry up to $12\%$ of the response, but none fire on most of the soft neutral particles without also firing on real JetClass particles, which the SAE latents do.\\
In the larger models, the SAE edits are less effective and the response is spread over more attention heads (Appendix~\ref{app:sizes}, Fig.~\ref{fig:app_readers}). We apply the same procedure to three readers in the medium model and four in the large one, which leave remaining responses of $0.35$ and $0.58$ (Appendix~\ref{app:mec-interp}).

\subsection{The weight-space programming approach}
\label{sec:wsp}

WSP is a recently introduced approach in which a large language model (LLM) directly inspects and modifies the weights of a neural network, without relying on gradient-based optimization~\cite{takehi2026weightspace}. Instead, it can use the model's equations and closed-form calculations on its weights and captured activations to reason about the model. We focus on how the information supplied to the agent affects the repair.

\begin{figure}[h!]
  \centering
  \includegraphics[width=0.6\textwidth]{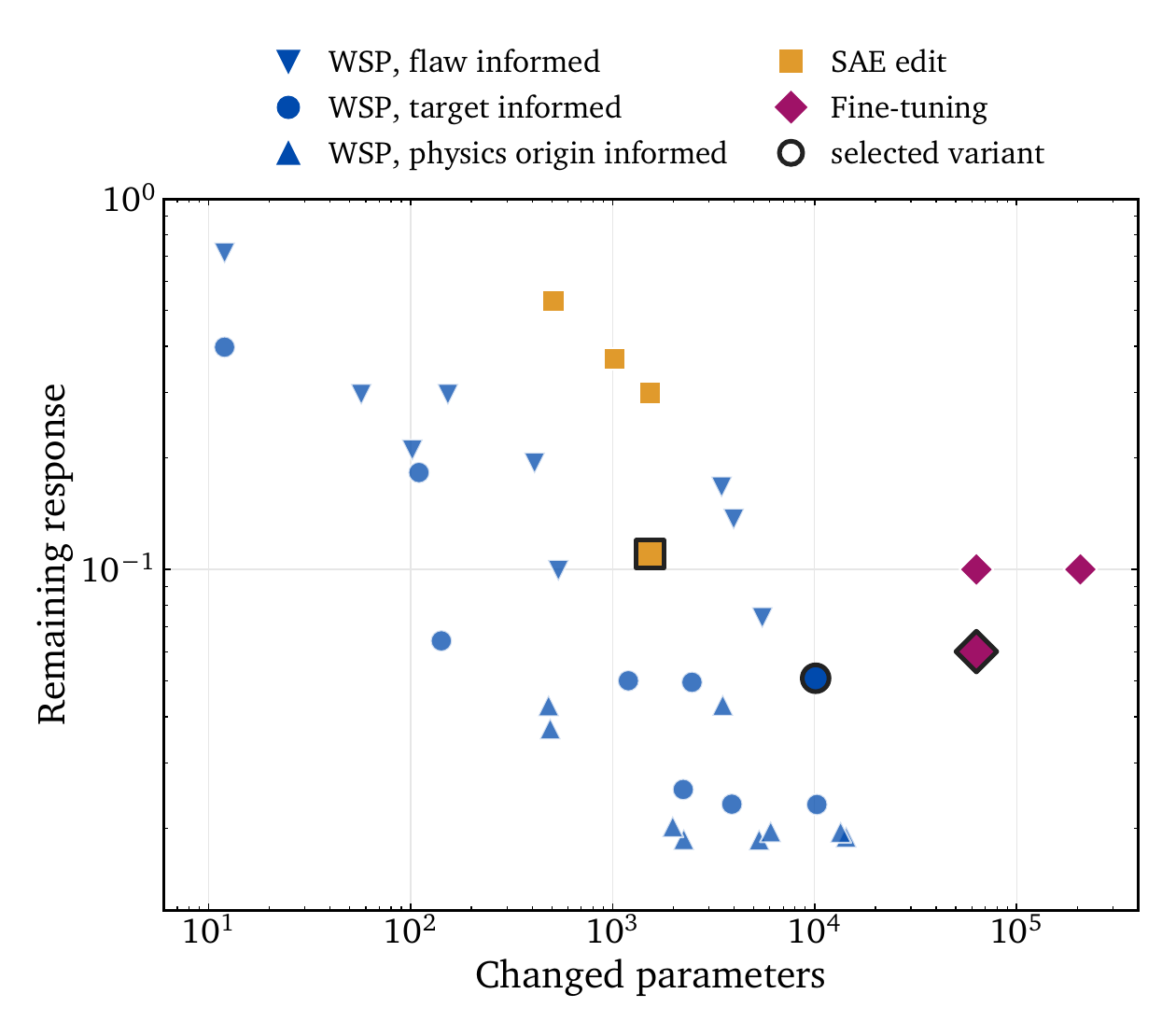}
  \caption{Remaining response against the number of changed parameters, for all the repair variants tested on the smallest model. The fine-tuning points are the variants of Tab.~\ref{tab:ft_variants} and the SAE points those of Tab.~\ref{tab:sae_variants}. Each WSP point is one edit of one run, with three edits per run and three runs per variant; Tab.~\ref{tab:wsp_variants} gives the median and range over the runs. The black outline marks the variant selected for each method (Tab.~\ref{tab:comparison}); for WSP it marks the median run of the selected variant, which is used in Sec.~\ref{sec:heads}.}
  \label{fig:frontier}
\end{figure}

The agents all use the same LLM, Claude Opus 5.5~\cite{anthropic2026claudeopus55}, run through the Claude Code agent, and receive the model code, the pretrained weights, and $10\,240$ practice jets. They cannot run the model themselves: each run allows up to $50$ queries to an evaluation tool, which returns either the model predictions, its internal activations, or the scores of a proposed edit. The agents can modify the body and classifier weights, while the generator weights remain fixed. They are not told where in the network the flaw is located.
\\
A simple way to remove the flaw outside the model is to drop every neutral particle below the JetClass reconstruction thresholds before the jet enters the network. We call this the removal rule. A repaired model should give, on the full jet, the predictions that the original model gives after the removal rule. We call these the target predictions, and we score all edits against them.
\\
We compare three variants, which differ only in the information given to the agent:
\begin{itemize}
    \item \textbf{Flaw informed} The agent is told that one very soft neutral particle changes the classification, and that the repaired model should behave like the original model on the jet without its very soft neutral particles. It is not told which particles count as very soft, and it cannot see the target predictions.
    \item \textbf{Target informed} The agent can ask for the target predictions on any jet, but it is told neither the flaw nor the removal rule. It must first find out where the target differs from the original model, and only then build a repair.
    \item \textbf{Physics origin informed} The agent is told the flaw, its origin in the dataset shortcut, the reconstruction thresholds, and the code of the removal rule. It only has to find how to write this known rule into the weights.
\end{itemize}
For each model size, we launch each variant in three independent runs, with the same prompt. Each run returns three sets of edited weights, which differ in the number of parameters they change. We evaluate all of them, and Fig.~\ref{fig:frontier} shows them all. In the tables, each run is represented by its edit with the lowest remaining error, defined as the distance to the target predictions divided by the same distance for the original model. Tab.~\ref{tab:wsp_variants} gives the median of these three edits, one per run, with the range in brackets.

\begin{table}[h]
  \centering
  \small
  \setlength{\tabcolsep}{4pt}
  \newcommand{\cell}[2]{\makecell{#1\\{\scriptsize[#2]}}}
  \begin{adjustbox}{max width=\textwidth}
  \begin{tabular}{@{}l ccccc@{}}
    \toprule
    Variant
      & \makecell{Remaining\\error}
      & \makecell{Remaining\\response}
      & \makecell{$\Delta$\,accuracy\\(clean)}
      & \makecell{Physical\\response}
      & \makecell{Changed\\parameters} \\
    \midrule
    Flaw informed
      & \cell{$0.33$}{0.26, 0.37}
      & \cell{$0.14$}{0.10, 0.21}
      & \cell{$-0.0003$}{$-$0.0021, 0.0000}
      & \cell{$0.95$}{0.89, 0.96}
      & \cell{$539$}{102, 3\,976} \\
    Target informed (selected)
      & \cell{$0.12$}{0.08, 0.23}
      & \cell{$0.05$}{0.02, 0.06}
      & \cell{$-0.0002$}{$-$0.0012, +0.0003}
      & \cell{\textbf{1.01}}{0.93, 1.01}
      & \cell{$10\,122$}{142, 10\,267} \\
    Physics origin informed
      & \cell{\textbf{0.08}}{0.06, 0.09}
      & \cell{\textbf{0.02}}{0.02, 0.04}
      & \cell{$-0.0002$}{$-$0.0004, +0.0002}
      & \cell{\textbf{0.99}}{0.99, 0.99}
      & \cell{$13\,456$}{491, 14\,321} \\
    \bottomrule
  \end{tabular}
  \end{adjustbox}
  \caption{Weight-space programming variants on the smallest model. Each cell gives the median over three runs and, below it, the range. Columns as in Tab.~\ref{tab:comparison}, with the remaining error in addition: the distance to the target predictions, divided by the same distance for the original model. The best median of each column is in bold.}
  \label{tab:wsp_variants}
\end{table}

More information increases the agents' performance. The remaining error of the least informed agents is $0.33$, against $0.12$ and $0.08$ for the two other variants, and the ranges do not overlap. They also change the response to physical changes more, and their edits are the smallest, with a few hundred parameters. The two variants that have access to the target are close to each other on all three quantities. The physics origin does not help the agent more than the target alone.
For this reason, we select the target informed variant. It reaches a repair close to the physics origin informed variant, without being given the removal rule or the origin of the flaw.
\\
However, the explanation makes the repair faster. After ten queries, the physics origin informed agent has a remaining error below $0.2$ on the practice jets in all three runs, and the target informed agent in one. The flaw informed agent is never shown its remaining error, so this comparison does not exist for it.

The three variants proceed differently. The flaw informed agents do not know which particles count as very soft, and have to choose a threshold themselves. The target informed agents add particles to jets and vary their momentum, energy, and type, which is enough for them to discover that the target ignores neutral particles below a threshold. The physics origin informed agents only have to write the given rule into the weights.
\\
Most repairs follow the same idea: a small detector for soft neutral particles, built from a few units of an MLP in an early block, followed by a change that hides the detected particles from the attention or makes the network treat them as padding. One run instead reuses a feature that already exists in a late layer.
\\
None of the edits acts mainly on \texttt{b7/6}, the main reader of Sec.~\ref{sec:IR_location}. The agents add a new detector upstream of the readers instead of changing the readers themselves.
\\
Each agent works in its own folder, and we record every file it opens and every command it runs. No agent read the files of another run or of our analysis. In two runs of the flaw informed variant, on the medium and large models, the agent listed the running processes and saw command lines of other agents working at the same time. We keep these runs and report them in Appendix~\ref{app:control}.

The weak point of the method is the variation between runs. The same prompt gives edits whose remaining error differs by up to a factor of three, and whose size ranges from $142$ to about $10\,000$ parameters. The effect on the other heads of the model also depends on the run, which we study in Sec.~\ref{sec:heads}. Within a run, a larger edit is closer to the target predictions. Across runs, the size alone does not predict the repair, and less so for the larger models: the $142$-parameter edit of one run has a remaining response of $0.06$, against $0.05$ for the $10\,122$-parameter edit of another (Fig.~\ref{fig:frontier}).
An edit is small enough to be inspected, but one run is not enough to trust the method.

The agents also repair the medium and large models: for the selected variant, the remaining error is $0.04$ and $0.10$, and the flaw informed agents again repair less, with $0.23$ and $0.32$. Details are given in Appendix~\ref{app:wsp} and Appendix~\ref{app:sizes}.

\subsection{Comparison of the approaches}
\label{sec:comparison}

We now compare the variant selected for each of the three methods in the previous subsections. All of them are evaluated on the same test jets, which none of the methods has used. The quantities are defined in Appendix~\ref{app:repair-method}. Tab.~\ref{tab:comparison} gives the results for the smallest model, and Fig.~\ref{fig:frontier} shows the repair against the size of the change for all the variants we tested.

\begin{table}[hb!]
  \centering
  \small
  \setlength{\tabcolsep}{3pt}
  \begin{adjustbox}{max width=\textwidth}
    \begin{tabular}{@{}l cc ccc c l@{}}
    \toprule
    & \multicolumn{2}{c}{\textbf{Repair}}
    & \multicolumn{3}{c}{\textbf{Impact on the model}}
    & \multicolumn{2}{c}{\textbf{Requirements}} \\
    \cmidrule(lr){2-3}\cmidrule(lr){4-6}\cmidrule(l){7-8}
    Method
      & \makecell{Remaining\\response}
      & \makecell{W eff.\\(+1 $\gamma$)}
      & \makecell{$\Delta$\,acc.\\(clean)}
      & \makecell{Physical\\response}
      & \makecell{Changed\\parameters}
      & \makecell{Gradient}
      & \makecell[l]{Prior knowledge} \\
    \midrule
    Original model      & $1$     & $54\%$ & --  & --  & --  & --  & -- \\
    Input removal rule  & $0.01$  & $79\%$ & $0$ & $1$ & $0$ & no  & thresholds \\
    \midrule
    Fine-tuning         & $0.06$        & \textbf{79\%} & $-0.0001$ & $0.90$        & $63\,240$       & yes      & loss, trained modules \\
    SAE edit            & $0.11$        & $78\%$        & $-0.0006$ & $0.78$        & \textbf{1\,533} & SAE only & reader locations \\
    WSP                 & \textbf{0.05} & \textbf{79\%} & $-0.0002$ & \textbf{1.01} & $10\,122$       & no       & target predictions \\
    \bottomrule
    \end{tabular}

  \end{adjustbox}
  \caption{Comparison of the repair methods on the smallest model, on held-out jets. \emph{Repair}: response of the JetClass outputs to one added photon or neutral hadron at or below the softest neutral particles in JetClass, relative to the original model, and $W$ efficiency at the $80\%$ working point after adding one photon at $f=10^{-3}$. \emph{Impact on the model}: change of the accuracy on unmodified jets, response of the JetClass outputs to physically meaningful changes of the jet relative to the original model, and number of free parameters of the weight change. \emph{Requirements}: whether gradients are used, and what must be supplied before the method can run. For WSP, the values are the median over the three runs of the selected variant; the ranges are given in Tab.~\ref{tab:wsp_variants}.
}
  \label{tab:comparison}
\end{table}

\paragraph{Repair}
All three approaches bring back this efficiency between $78\%$ to $79\%$ while the efficiency of the original model falls from $80\%$ to $54\%$.
The SAE edit removes most of the flaw but its remaining response is $0.11$ against $0.05$ and $0.06$ for the others.

\paragraph{Impact on the model}
WSP is the method that changes the rest of the model the least. It keeps the response to physical changes within $1\%$ of the original, against $0.90$ for fine-tuning and $0.78$ for the SAE edit. It changes about $10\,000$ parameters, six times fewer than fine-tuning. The SAE edit is the smallest, with $1\,533$ parameters. No method changes the accuracy on unmodified jets. The effect on the generator head is studied in Sec.~\ref{sec:heads}.

\paragraph{Requirements}
The methods differ in what must be known before they can run. Fine-tuning needs a loss built by hand and a choice of the weights to train. The SAE edit needs the location of the readers from Sec.~\ref{sec:IR_location}. WSP uses no gradients and needs only the target predictions; without them, the agents repair much less (Sec.~\ref{sec:wsp}).

\paragraph{Limits}
The WSP results vary between runs (Sec.~\ref{sec:wsp}), and for the large model, its edits reach up to $400\,000$ parameters (Appendix~\ref{app:sizes}). The agents were also strongly constrained for this study, which we discuss in Sec.~\ref{sec:outlook}.

\section{Testing the repaired OmniLearned model}
\label{sec:repaired}

We initially found the infrared sensitivity on the $W$ tagger, while building a control for an interpretability study. Also, all three repairs of the previous section are built using only the classifier outputs for JetClass jets. A repair could in principle only suppress the response of these outputs and leave the sensitivity everywhere else. If it instead targets the cause found in Sec.~\ref{sec:hypothesis} that jets with soft particles are identified as JetClass-II jets by the backbone, it should also repair every head and task of the model that relies on this latent representation of the backbone. We test the representation itself (Sec.~\ref{sec:representation}), the taggers of the classification head (Sec.~\ref{sec:taggers}), and the generator head and a new regression head (Sec.~\ref{sec:heads}).

\subsection{The repaired representation}
\label{sec:representation}

We now compare the representation itself, before and after the repair. We use the four summary tokens after the last block of the body, which is the input of the classifier head, and compare the original model with the model repaired by the selected WSP edit.

\begin{figure}[ht]
  \centering
    \includegraphics[width=0.9\textwidth]{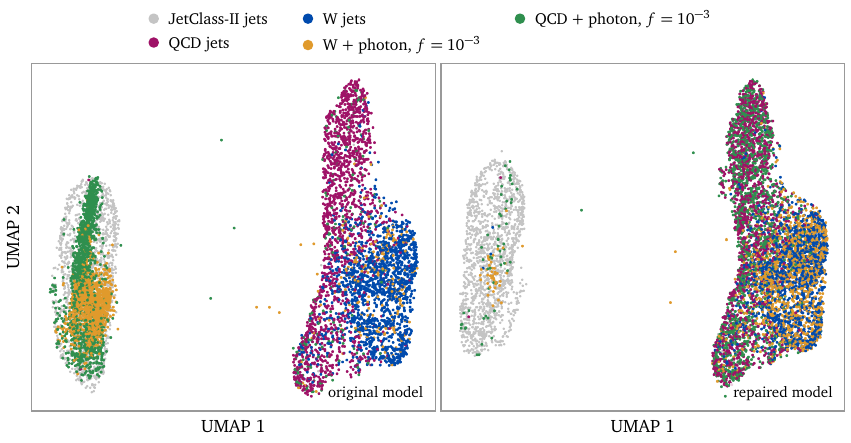}
  \caption{UMAP projection of the input tokens of the classifier, for the original model (left) and the repaired model (right). The projection is fitted on the unmodified jets of each model separately, so the two panels do not share their axes.}
  \label{fig:latent}
\end{figure}

Fig.~\ref{fig:latent} shows a two-dimensional projection of these tokens. In the original model, JetClass and JetClass-II jets form two separate groups, and the JetClass jets with one added soft photon lie in the JetClass-II group. After the repair, the two groups are still separate, but the jets with a soft photon lie with the $W$ and QCD jets they come from. For a $W$ jet with one added soft photon, $86\%$ of its $15$ nearest neighbors are JetClass-II jets in the original model compared to only $3\%$ in the repaired one.

\subsection{The taggers}
\label{sec:taggers}
Using each of the nine JetClass signal classes OmniLearned is pretrained on, we can define a tagger that can be used without fine-tuning, in the same way as the $W$ tagger of Sec.~\ref{sec:wtag}. We found in Sec.~\ref{sec:hypothesis} that all nine taggers are infrared unsafe. Table~\ref{tab:taggers} shows their signal efficiency after one soft photon is added. In the model repaired by WSP they stay between $78\%$ and $80\%$ for all nine taggers. On unmodified jets the repair does not change the efficiency. Fine-tuning and the SAE edit also restore all nine taggers, to $74\%$ or more.

\begin{table}[h!]
\centering
\setlength{\tabcolsep}{4pt}
\begin{adjustbox}{max width=\textwidth}
\begin{tabular}{lccccccccc}
\toprule
&$H{\to}b\bar b$ & $H{\to}c\bar c$ & $H{\to}gg$ & $H{\to}4q$ & $H{\to}\ell\nu qq$ & $Z{\to}q\bar q$ & $W{\to}qq$ & $t{\to}bqq$ & $t{\to}b\ell\nu$
 \\
\midrule
Original & 37 & 57 & 42 & 50 & 12 & 52 & 54 & 61 & 13 \\
Repaired & 79 & 80 & 79 & 79 & 78 & 79 & 79 & 80 & 78 \\
\bottomrule
\end{tabular}
\end{adjustbox}
\caption{Signal efficiency (in \%) of the small model's nine JetClass taggers after one soft photon is added to the jet, for the original model and the model repaired by WSP. Each tagger's working point is at $80\%$ efficiency.}
\label{tab:taggers}
\end{table}

\begin{figure}[h!]
  \centering
  \includegraphics[width=\textwidth]{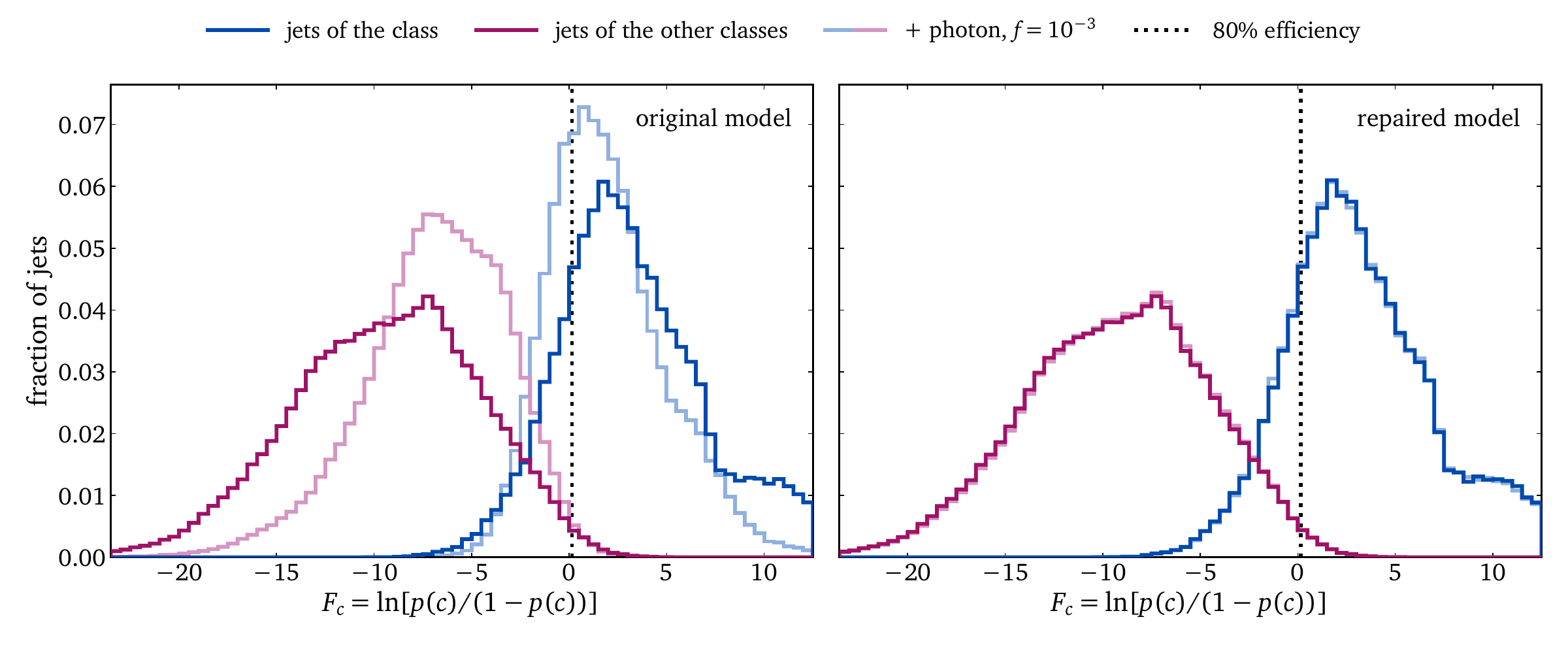}
  \caption{Distribution of the one-vs-rest score $F_c$ of the ten JetClass classes, for the jets of class $c$ and for the jets of the nine other classes, summed over the ten classes, in the original model (left) and in the model repaired by WSP (right). The light lines show the same jets with one added photon at $f=10^{-3}$. The dotted line is the single cut that keeps $80\%$ of the unmodified jets of class $c$.}
  \label{fig:taggers}
\end{figure}

The same holds when each class is separated from all the other JetClass classes instead of from QCD only. Fig.~\ref{fig:taggers} shows the distribution of the one-vs-rest score $F_c=\ln[p(c)/(1-p(c))]$, where $p(c)$ is the probability of class $c$ among the ten JetClass classes, for the jets of the class and for the jets of the other classes, summed over the ten classes. In the original model one soft photon shifts both distributions, and the efficiency of a cut that keeps $80\%$ of the unmodified jets of the class falls to $70\%$ once the photon is added. In the repaired model the distributions with and without the photon coincide, and the efficiency stays at $80\%$.

\paragraph{The response to softer particles}
\label{sec:softer}

We test whether the three approaches make OmniLearned more infrared safe for particles much softer than those it was built with. Down to $1~\mev$, the response of the three repaired models to an added photon or neutral hadron stays flat and at most at the same response level as an added charged hadron, while the response of the original model remains high throughout (Appendix~\ref{app:softer}). No response vanishes completely as $\pT\rightarrow0$, so the conservative claim is that the repairs remove the excess response to soft neutral particles that comes from the dataset shortcut, but none of them makes the tagger infrared safe in the formal sense.

\subsection{Robustness across heads}
\label{sec:heads}

All heads share their body with the classifier, so the flaw may reach them too. While the representation at the input of the heads appears repaired (Sec.~\ref{sec:representation}), previous work shows that heads can use the same latent differently~\cite{BonnetGuerrini2026FindingAU}. The other heads thus offer a natural independent test of the robustness of the fix.

\paragraph{The generator head}

The generator is trained with flow matching~\cite{Lipman2022FlowMF}: starting from noise, it predicts at each step how the momentum of each particle has to be updated to move the noised jet toward a real jet. We take $W$ jets with little noise ($t=0.05$), add one photon of $0.3$~GeV, below the JetClass threshold, run the generation to the end, and measure the relative change of the mass of the generated jet. As a control, we use a photon of $3$~GeV, above the threshold.\\
In the original model the sub-threshold photon changes the mass of the generated jet by $1.1\%$, almost three times as much as the ten times harder control photon ($0.4\%$), so the flaw reaches the generator. All three repairs bring this response down to the level of the control or below, with medians of $0.2\%$ for WSP and fine-tuning and $0.3\%$ for the SAE edit (Appendix~\ref{app:generator}).

\paragraph{The regression head}

In Sec.~\ref{sec:hypothesis} a jet mass regression head trained on the frozen backbone inherited the flaw and one soft photon changed its prediction by $10.6~\gev$. We repeat this test on the repaired models, training a new head on the output of the backbone of each of the repair variants. On the models repaired by WSP and SAE edits, the prediction changes less ($1.1~\gev$) than for a control photon of $3~\gev$, above the threshold ($2~\gev$). For these two repairs a new head no longer responds to the sub-threshold photon more than to one above the threshold, without any further correction, which shows that the two approaches are not just fixing the symptom, but repairing the cause. With fine-tuning about half of the response remains ($5.0~\gev$ $[4.7, 5.6]$ over the three seeds).

\section{Outlook}
\label{sec:outlook}
Our results show that sparse-autoencoder-guided interventions and weight-space programming can correct unwanted behavior in a particle-physics foundation model through direct edits to its weights. SAE-guided edits target the latent directions read by specific attention heads, while language-model agents construct corrections without gradient-based optimization. On the smallest model, WSP achieves a correction comparable to fine-tuning with fewer free parameters and better preservation of the tested physical responses, while for the SAE approach removing three latents is sufficient. These approaches, WSP and SAE-guided edits in particular, offer a way to revise expensive pretrained models while retaining their useful capabilities.

Beyond the potential to reduce computational costs, surgical interventions can be inspected, linked to a proposed mechanism, and tested individually. For SAE-guided edits, we identify the feature being suppressed and directly control the strength and location of the intervention.

Mechanistic interpretability and weight-space programming play complementary roles in this process. In OmniLearned, mechanistic analysis identifies a dataset-provenance shortcut responsible for the sensitivity to soft neutral particles and enables targeted SAE-guided edits. Our comparison shows that agents given access to predictions of this desired behavior achieve better corrections than those given only a description of the flaw. Mechanistic analysis can therefore guide both where to intervene and what an intervention should achieve.

After our repairs, the response to a very soft particle still leaves nonzero values. This suggests that, to guarantee complete infrared safety, it has to be built into the architecture, or that it might only emerge as a condition in physics-aware models.

The WSP results were obtained under tightly constrained conditions, using an LLM that was not state of the art, with at most 50 evaluation queries per run and edits restricted to the backbone and classification head. This leaves open how repair quality and consistency scale with agent capability, evaluation budget, and access to mechanistic information. The transfer of corrections to other signal classes and a newly trained regression head encourages studying these methods beyond the task used to develop the repair. Variation across agent runs and in the generator response also shows that the effects of an edit must be assessed across downstream tasks.

More broadly, direct weight editing could allow physics requirements identified after pretraining to be incorporated into existing models (see Authors' position note). Future work could investigate objectives that are difficult to achieve through conventional fine-tuning and test whether mechanistically informed edits can improve predictive performance or introduce new capabilities, extending their role beyond correcting unwanted behavior.

\section*{Authors' position note}
\label{sec:authors}
WSP remains a recent approach, and its scope and implications are still unclear. We believe that such a powerful tool should be used with great care. We personally support its usage in only a restricted, controlled manner.
\\
While it offers opportunities to improve the interpretability and alignment of large models, it would be naive to overlook its potential misuse. Direct access to model weights could allow users to introduce intentional biases or weaken safeguards. It could also contribute to systems that modify their own behavior, performing autonomous self-improvement. Although our study does not demonstrate any of these possibilities, they call for careful consideration as the method develops.

\section*{Code and data availability}
\label{sec:availability}
The code used for the analysis, the fine-tuning, the SAE edits and the scoring is available in this \href{https://github.com/rbonnetguerrini/WSP-mech-interp-for-model-editing-in-particle-physics}{GitHub repository}, together with the three task descriptions with their checksums, the edits handed in by the agents with their scores, the notes and full transcripts of the agents, and the access records. The edited weights of all the variants, the SAE checkpoints and the jet sets are archived on Zenodo, \href{https://doi.org/10.5281/zenodo.24377675}{10.5281/zenodo.24377675}.

\section*{Acknowledgments}
We thank the participant at the IML conference for the question regarding neutral hadrons at a presentation of an early result of the study. We thank Carlos Moreno Martinez for comments and discussion on an initial draft of the paper. We thank the authors of OmniLearned for making OmniLearned's weights publicly available. This work was supported by the European Union's Horizon Europe research and innovation programme under the Marie Skłodowska-Curie grant agreement No. 101168829, Challenging AI with Challenges from Physics: How to solve fundamental problems in Physics by AI and vice versa (AIPHY). We thank the CINECA Leonardo HPC cluster (via INFN grant PML4HEP) and the University of Copenhagen's SCIENCE AI Centre for providing computing resources.

\clearpage
\bibliographystyle{UTPstyle}
\bibliography{ref}
\clearpage
\appendix

\phantomsection
\addcontentsline{toc}{section}{\protect\numberline{}Appendix}
\section*{Appendix}

\addtocontents{toc}{\protect\setcounter{tocdepth}{0}}

\section{Details on the response of OmniLearned to soft particles}
\label{app:figs}

\begin{figure}[ht]
  \centering
  \begin{minipage}{0.49\textwidth}\centering
    \includegraphics[width=\textwidth]{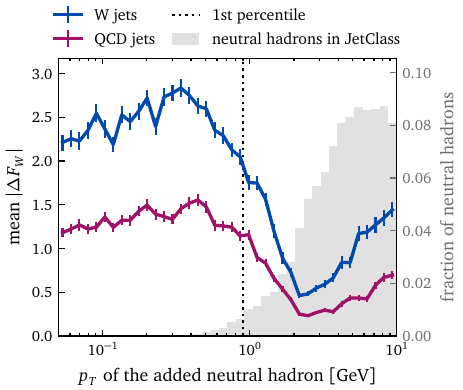}
  \end{minipage}\hfill
  \begin{minipage}{0.49\textwidth}\centering
    \includegraphics[width=\textwidth]{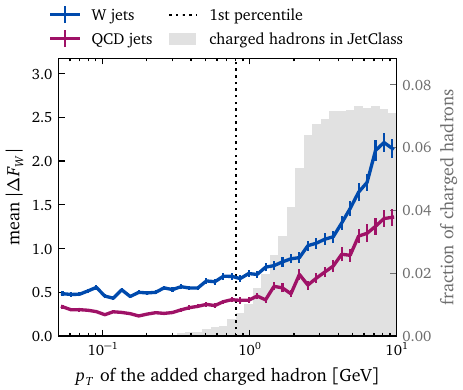}
  \end{minipage}
  \caption{Mean change of $F_W$ for one added neutral hadron (left) and one added charged hadron (right), as a function of its $\pT$, for $W$ and QCD jets. The gray histogram is the $\pT$ distribution of the respective particles in JetClass jets.}
  \label{fig:app_types}
\end{figure}

\begin{figure}[tbp]
  \centering
  \begin{minipage}{0.49\textwidth}\centering
    \includegraphics[width=\textwidth]{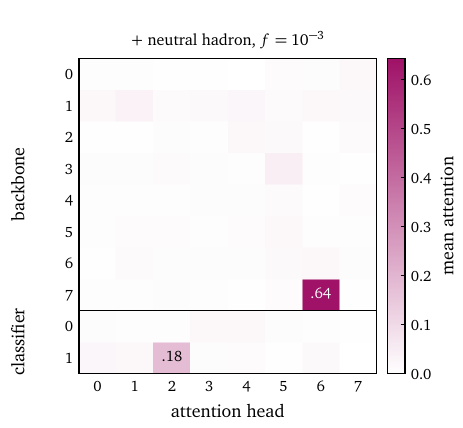}
  \end{minipage}\hfill
  \begin{minipage}{0.49\textwidth}\centering
    \includegraphics[width=\textwidth]{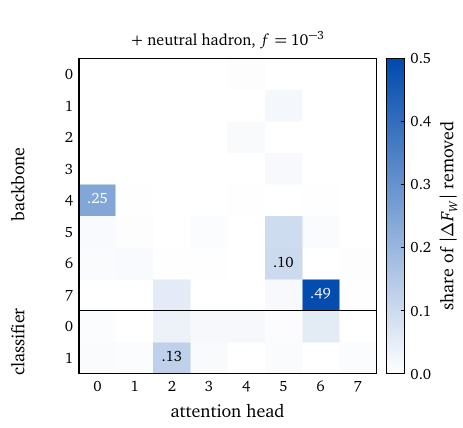}
  \end{minipage}\\[1.5ex]
  \begin{minipage}{0.55\textwidth}\centering
    \includegraphics[width=\textwidth]{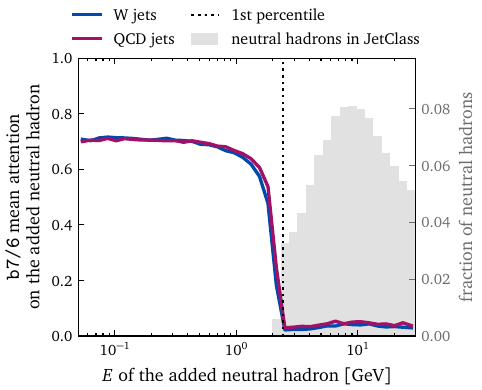}
  \end{minipage}
  \caption{Top left: mean attention of the four summary tokens on one added neutral hadron at $f=10^{-3}$, for every head of the OmniLearned backbone and classifier. Top right: share of $|\Delta F_W|$ removed when the output of one head is restored to its value on the clean jet, for the same neutral hadron. \texttt{b7/6} and \texttt{c1/2} attend strongly to the particle through their summary tokens, while \texttt{b4/0} reads it through the particle tokens. Bottom: mean attention of \texttt{b7/6} on one added neutral hadron as a function of its energy, for $W$ and QCD jets. The gray histogram is the energy distribution of neutral hadrons in JetClass jets; the dotted line marks the first percentile.}
  \label{fig:attention}
\end{figure}

This appendix contains additional figures of the analysis of Sec.~\ref{sec:wtag}.
Fig.~\ref{fig:app_types} shows the $W$ tagger's $\pT$ scan for one added neutral hadron and one added charged hadron (left). For neutral hadrons, the response is close to the response of photons (Fig.~\ref{fig:hook}, right). The step is broader in $\pT$ than for photons, because the JetClass threshold for neutral hadrons lies in energy (Fig.~\ref{fig:attention}, bottom). Charged hadrons on the other hand result in a flat response of the classification score  and their response rises smoothly with $\pT$ for photons in the training dataset. This is consistent with the reader heads responding only to neutral particles (Sec.~\ref{sec:IR_location}).

Fig.~\ref{fig:attention} (top left) shows how much the four summary tokens attend to a neutral hadron at $f=10^{-3}$, for each head of the backbone. The response is very close to that reported in Sec.~\ref{sec:IR_location} for neutral hadrons. In the bottom panel of Fig.~\ref{fig:attention} we show the attention of \texttt{b7/6} as a function of the energy of the added neutral hadron. Compared to the $\pT$ response, shown in Fig.~\ref{fig:app_types} (left), the response is much sharper in energy, and independent of pseudorapidity. This shows that OmniLearned learns an energy threshold for soft neutral hadrons. In the top right panel of Fig.~\ref{fig:attention}, we pass the jet with the added neutral hadron through the network, but replace the output of one head by its value on the clean jet. We then measure which share of the mean $|\Delta F_W|$ this removes, on 400 $W$ and QCD jets respectively. The replacement is applied at the summary tokens and at all particles that are unchanged, so it isolates what the head passes on from the added particle. As discussed in Sec.~\ref{sec:IR_location}, heads \texttt{b7/6}, \texttt{b4/0}, \texttt{c1/2} remove most of the response of the $W$ tagger, also for soft neutral hadrons. No other head removes more than about $10\%$. The summary tokens of \texttt{b4/0} hardly attend to the added particle, so the effect they carry reaches them through the particle tokens.

\begin{figure}[h]
  \centering
  \includegraphics[width=0.65\textwidth]{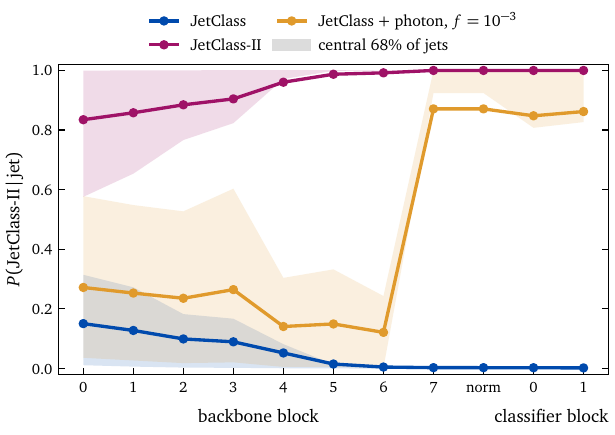}
  \caption{Probability that a jet comes from JetClass-II using a linear probe on the summary tokens after each transformer block of the shared backbone (b0-b7), the final normalization and the two classifier blocks (c0, c1). The probe is trained on clean JetClass and JetClass-II jets and applied to held-out jets. The band indicates the central $68\%$ probabilities for the different populations.}
  \label{fig:probes}
\end{figure}

For Fig.~\ref{fig:probes} we train one logistic-regression probe per layer on the four concatenated summary tokens using $2000$ clean $W$ and QCD JetClass jets and $2000$ clean $W$ and QCD JetClass-II jets, and apply it to $1000$ held-out jets per set. The probe can separate the two datasets increasingly well with depth. JetClass jets with one photon at $0.3~\gev$ stay between the two up to the sixth transformer block of the backbone, after which they are indistinguishable from JetClass-II jets. The central $68\%$ have a JetClass-II probability above $0.997$, so the reassignment happens not only on average, but for almost every jet.

\section{Details on the repairing methodology}
\label{app:repair-method}
Across Sec.~\ref{sec:comparison}, all quantities are computed on $20\,480$ JetClass test jets that no method has used. They use the centered JetClass logits $c_\theta$ of Eq.~\eqref{eq:ft_centred} and the scale $S_0$ of Eq.~\eqref{eq:ft_scale}; $\theta_0$ denotes the original model.

\textbf{Remaining response} tells how much of the model's reaction to a soft neutral particle is left after the repair.
For a modification $a$ that adds one soft particle to a jet, the response of a model is
\begin{equation}
    r_a(\theta) = \frac{1}{S_0}\,\mathbb{E}_x\!\left[\left\|c_\theta(a(x))-c_\theta(x)\right\|_2\right].
\end{equation}
The remaining response is the mean of $r_a(\theta)/r_a(\theta_0)$ over four modifications: one photon or one neutral hadron, at $f=10^{-3}$ or $f=10^{-4}$. It is $1$ for the original model and $0$ for a model that ignores these particles.

\textbf{W efficiency} tells what the repair means for a physics selection: a repaired tagger should keep $80\%$ of the $W$ jets, whether or not a soft photon is present.
\\
We fix the threshold on $F_W$ that keeps $80\%$ of the $W$ jets for the model under test, and give the fraction of $W$ jets that still pass it after one photon at $f=10^{-3}$ is added.

\textbf{Accuracy} tells whether the repair has changed the classification of ordinary jets.
$\Delta$\,acc.\ is the change of the ten-class JetClass accuracy on unmodified jets.

\textbf{Response to physical changes} tells whether the model still reacts to real changes of the jet as it did before: a value of one means the reaction is unchanged, a value below one that the model has become less sensitive to the physics.
For each of the three transformations $T$ of Eq.~\eqref{eq:ft_phys}, we compute the ratio
\begin{equation}
    \rho_T = \frac{\mathbb{E}_x\!\left[\left\|\Delta_T c_\theta(x)\right\|_2\right]}{\mathbb{E}_x\!\left[\left\|\Delta_T c_{\theta_0}(x)\right\|_2\right]}.
\end{equation}
For the change of fragmentation, particles below the JetClass thresholds are removed before the model is applied. We report the $\rho_T$ that is furthest from one.

\textbf{Changed parameters} tells how large the modification of the model is, independently of how it was obtained.
For a weight matrix of shape $m\times n$ whose change has rank $r$, we count $\min\!\left(N,\; r(m+n-r)\right)$ free parameters, where $N$ is the number of changed entries. For a bias or normalization vector, we count the changed entries. The total is the sum over all changed tensors.

\textbf{Remaining error} tells how far the edited model is from the model we asked for, on all jets and not only on those with an added soft particle.
For WSP, we also measure the distance to the target predictions of Sec.~\ref{sec:wsp}. For a set of modified jets $x_v$,
\begin{equation}
    e_v(\theta) = \frac{1}{S_0}\,\mathbb{E}_x\!\left[\left\|c_\theta(x_v)-c_{\mathrm{target}}(x_v)\right\|_2\right],
\end{equation}
and the remaining error is the mean of $e_v(\theta)/e_v(\theta_0)$ over the sets of modified jets. It is $1$ for the original model and $0$ for a perfect repair.

\subsection{Fine-tuning}
\label{app:fine-tuning}
We detail the training objective, the choice of parameters to update, and the tests that motivate the selected setup.

\paragraph{Training objective.}
Let $y_\theta(x)$ denote the 210 classifier logits of the model with weights $\theta$, and let $\theta_0$ denote the pretrained weights. A common shift of all ten JetClass logits leaves their relative probabilities, and therefore the W-versus-QCD score, unchanged. We remove this common shift so that the repair loss penalizes changes in the model's preference between classes. Writing $\mathcal{J}$ for the set of ten JetClass indices, we define
\begin{equation}
    c_{\theta,j}(x)
    = y_{\theta,j}(x)
    - \frac{1}{10}\sum_{k\in\mathcal{J}}y_{\theta,k}(x),
    \qquad j\in\mathcal{J}.
    \label{eq:ft_centred}
\end{equation}

We measure changes of these centered logits relative to the typical difference between two independent clean jets, $x$ and $x'$.
This defines a fixed scale from the pretrained model:
\begin{equation}
    S_0^2
    = \mathbb{E}_{x,x'}\!\left[
        \left\|c_{\theta_0}(x)-c_{\theta_0}(x')\right\|_2^2
      \right].
    \label{eq:ft_scale}
\end{equation}
We estimate this scale on the training sample. The expectations below average over the training jets and, where applicable, the random input modifications. The three terms in Eq.~\eqref{eq:ft_loss} are defined below.
\begin{itemize}
    \item $\mathcal{L}_{\mathrm{hold}}$ penalizes changes to the pretrained model's predictions on unmodified jets:
    \begin{equation}
        \mathcal{L}_{\mathrm{hold}}
        =
        \mathbb{E}_x\!\left[
            \frac{
                \sum_{k=1}^{210}
                w_k\left(y_{\theta,k}(x)-y_{\theta_0,k}(x)\right)^2}
                {\sum_{k=1}^{210}w_k}
          \right],
        \qquad
        w_k =
        \begin{cases}
            1, & k\in\mathcal{J},\\
            0.1, & k\notin\mathcal{J}.
        \end{cases}
        \label{eq:ft_keep}
    \end{equation}
        For the same change in one output, this term penalizes a JetClass logit ten times more strongly than another logit. We use uncentered logits here to also preserve their values relative to the other classes.
    \item $\mathcal{L}_{\mathrm{IC}}$ penalizes changes under soft-particle additions:
    \begin{equation}
        \mathcal{L}_{\mathrm{IC}}
        = \mathbb{E}_{x,\mathrm{aug}}\!\left[
            \frac{1}{|\mathcal{A}|}
            \sum_{a\in\mathcal{A}}
            \frac{
                \left\|c_\theta(a(x))-c_\theta(x)\right\|_2^2
            }{S_0^2}
          \right].
        \label{eq:ft_ic}
    \end{equation}
    Here $\mathcal{A}$ contains two modifications. The first adds one photon or neutral hadron, with equal probability, at a random position within $\Delta R<0.4$. Its transverse-momentum fraction is drawn uniformly in $\log f$ between $10^{-4}$ and $10^{-2}$. The second splits the hardest neutral particle into two collinear particles with equal momenta. We draw a new soft particle at each training step. The numerator measures the change caused by each modification. Dividing by $S_0^2$ expresses it relative to the variation between clean jets.  The collinear term targets an additional sensitivity of the model. The fine-tuning objective therefore extends beyond the soft-neutral threshold effect studied in Sec.~\ref{sec:irc}.
    \item $\mathcal{L}_{\mathrm{phys}}$ preserves the response to other changes in the jet. For a transformation $T$, we define
    \begin{equation}
        \Delta_T c_\theta(x)
        = c_\theta(Tx)-c_\theta(x).
    \end{equation}
    The loss compares this response with that of the pretrained model:
    \begin{equation}
        \mathcal{L}_{\mathrm{phys}}
        = \mathbb{E}_x\!\left[
            \frac{1}{|\mathcal{T}|}
            \sum_{T\in\mathcal{T}}
            \frac{
                \left\|
                    \Delta_T c_\theta(x)
                    - \Delta_T c_{\theta_0}(x)
                \right\|_2^2
            }{S_0^2}
          \right].
        \label{eq:ft_phys}
    \end{equation}
    Here $\mathcal{T}$ contains three changes: scaling all angles to the jet axis by $1.105$, scaling all momenta and energies by $1.05$, and making the fragmentation softer at fixed jet transverse momentum. We require the repaired model to respond to these changes in much the same way as the original model.  The fragmentation change can move particles below the reconstruction thresholds. During training, we keep these particles in the input. This can also trigger the soft-neutral sensitivity we aim to remove. We therefore perform an additional evaluation in which we remove particles below the thresholds. This tests whether the repair preserves the response to fragmentation changes among particles that JetClass would reconstruct.

\end{itemize}

The three loss terms serve complementary roles.
$\mathcal{L}_{\mathrm{hold}}$ preserves predictions on unmodified jets. $\mathcal{L}_{\mathrm{IC}}$ reduces changes caused by soft-particle additions and collinear splittings. $\mathcal{L}_{\mathrm{phys}}$ preserves the response to other changes in the jet.

\paragraph{Update scope.}
We allow updates in the body's attention modules and the network that computes the pairwise attention bias. This includes attention modules throughout the body, rather than only the identified heads. The input embedding, the remaining MLPs, the normalization layers, and both output heads remain frozen. The generator can still change its predictions through the shared body, so we evaluate generation separately.
\\
For each selected weight matrix, we use a low-rank update,
\begin{equation}
    W = W_0 + AB,
    \qquad \operatorname{rank}(AB)\leq 8,
    \label{eq:ft_lora}
\end{equation}
where the pretrained matrix $W_0$ remains fixed and only $A$ and $B$ are trained. We initialize $A$ to zero, so training starts from the original model. We also train additive changes to the two bias vectors in the pairwise interaction network. For the smallest model, this gives 22 low-rank matrix updates and two bias updates, with about $64\,600$ trainable parameters. The resulting weight change has $63\,240$ free parameters, which is the number we report in the tables. The rank constraint limits the number of free parameters; it does not directly limit the magnitude of the weight changes.
We train on $10\,240$ jets for $3000$ steps, with batches of 128. We use Adam with a learning rate of $10^{-3}$, a 100-step warm-up, and cosine decay. The loss coefficients are $\lambda=30$ and $\mu=10$.

\paragraph{Further tests.}

Tab.~\ref{tab:ft_variants} compares the selected setup with two alternatives: training on photons only, and updating the whole body and classifier.

\begin{table}[h]
  \centering
  \small
  \begin{tabular}{@{}l ccccc@{}}
    \toprule
    Variant
      & \makecell{Remaining\\response}
      & \makecell{$\Delta$\,accuracy\\(clean)}
      & \makecell{Physical\\response}
      & \makecell{Changed\\parameters} \\
    \midrule
    Attention only, both neutral types (selected) & \textbf{0.06} & \textbf{$-$0.0001} & $0.90$         & \textbf{63\,240} \\
    Attention only, photons only                  & $0.10$        & \textbf{$-$0.0003} & \textbf{0.92}        & \textbf{63\,240} \\

    Whole body and classifier, both neutral types & $0.10$        & $-0.006$           & $0.42$           & $206\,868$ \\
    \bottomrule
  \end{tabular}
  \caption{Fine-tuning variants on the smallest model. Columns as in Tab.~\ref{tab:comparison}. The best value of each column is in bold.}
  \label{tab:ft_variants}
\end{table}

The selected variant is trained with three random seeds, the other variants once. The three seeds give a remaining response of $0.061\pm0.002$ and a response to physical changes of $0.906\pm0.003$. For the choice of particles (Sec.~\ref{sec:fine-tuning}), adding neutral hadrons to the objective reduces the mean $|\Delta F_W|$ on $W$ jets from $0.73$ to $0.15$ for added neutral hadrons with $\pT$ between $0.85$ and $1.6~\gev$. Giving each training jet both a photon and a neutral hadron, instead of one of the two at random, gives similar results. Two earlier tests, on photons only and with updates across the body and classifier, motivated the other choices: without $\mathcal{L}_{\mathrm{phys}}$, the response to a change of fragmentation fell to $0.26$ of the original, and rank-1 updates gave a weaker repair than rank 8.

For the medium and large models, the selected recipe changes about $355\,000$ and $1.6$ million parameters. It leaves a remaining response of $0.03$ and $0.05$, with a response to physical changes of $0.95$ and $0.98$. After one added soft photon, the $W$ efficiency is $80\%$ for both, against $28\%$ and $40\%$ for the original models.

\subsection{Mechanistic Interpretability}
\label{app:mec-interp}
\paragraph{SAE training}
We train one TopK sparse autoencoder for each of the three readers, on the normalized particle-token activations entering the attention block that contains the reader. For the reader in the classifier head, these are the activations at the output of the body.
The training sample consists of $2.5$ million tokens from JetClass and \texttt{JetClass-II} jets. We center the activations and divide them by their root-mean-square distance from the mean. The SAE encodes each activation into a dictionary of $4096$ latents and retains only the $32$ largest positive activations per token. A decoder reconstructs the input from this sparse representation. We minimize the reconstruction error normalized by the total activation variance:
\begin{equation}
    \mathcal{L}_{\mathrm{SAE}}
    =
    \frac{\mathbb{E}_{h}\!\left[\|h-\hat{h}\|_2^2\right]}
    {\mathbb{E}_{h}\!\left[\|h-\mathbb{E}[h]\|_2^2\right]},
\end{equation}
where $h$ is the SAE input and $\hat{h}$ its reconstruction.
On a separate sample, the SAE trained at \texttt{b7/6} explains $95\%$ of the activation variance.

\paragraph{Latent identification} We search for latents that respond to neutral particles below the JetClass reconstruction thresholds. A latent fires when its activation is positive. We define its selectivity as its average firing rate on added photons at $\pT=0.3~\gev$ and neutral hadrons at $\pT=0.5~\gev$ (the two test particles of Sec.~\ref{sec:IR_location}), minus its firing rate on real JetClass neutral particles. We require an average firing rate of at least $50\%$ on the added particles and at most $5\%$ on real JetClass neutrals. The selectivity must also exceed the 99th percentile obtained with a frozen-random dictionary of the same size. We rank latents by selectivity. Among the 15 highest-ranked latents, those within 0.02 of the highest selectivity are distinguished by their decoder direction's alignment with the selected head's key readout: we select the one with the largest contribution to the attention logit for the mean clean summary-token query. We also measure its firing rate on sub-threshold neutral particles from \texttt{JetClass-II}. We apply the same procedure at the three readers, and select one latent at each.

\paragraph{Intervention}
To test the role of the identified latent, we subtract its decoded contribution from the added particle's original activation:

\begin{equation}
    h' = h-z_\ell(h)d_\ell,
\end{equation}

where $z_\ell(h)$ is the selected latent's activation, computed with the training normalization, and $d_\ell$ is its decoder direction expressed in the original activation units. We subtract this contribution from the original activation rather than replacing it with an SAE reconstruction. This preserves the reconstruction residual. The intervention acts at the attention block's shared input, rather than exclusively on the selected head. For an added photon at $0.3~\gev$, this lowers the attention of the head on the particle from $0.72$ to $0.09$ and reduces $|\Delta F_W|$ by $40\%$, close to the $49\%$ obtained by patching the head output for the same photon (Sec.~\ref{sec:IR_location} gives $58\%$ for a photon at $f=10^{-3}$). Removing another randomly selected active latent changes neither (attention $0.72$, $|\Delta F_W|$ reduced by less than $1\%$).

To obtain a repair that does not require the SAE while the model runs, we write the same operation into the weights. Let $u$ be the unit decoder direction of the latent and $W_K$ the weights with which the head computes its attention keys. We replace
\begin{equation}
    W_K \;\rightarrow\; W_K\left(I-\alpha\,u u^{\top}\right),
\end{equation}
with $\alpha=1$, or $\alpha=0.5$ for the half-strength variant. The head can then no longer distinguish particles along $u$. This is a rank-one change, with $511$ free parameters per reader for the smallest model. The selected edit applies this change to the three readers, with the latent of each, for $1\,533$ free parameters in total.

We compare edits of one, two and three readers, and of three readers at half strength in Tab.~\ref{tab:sae_variants}.

\begin{table}[h]
  \centering
  \small
  \begin{tabular}{@{}l ccccc@{}}
    \toprule
    Variant
             &\makecell{Remaining\\response}
      & \makecell{$\Delta$\,accuracy\\(clean)}
      & \makecell{Physical\\response}
      & \makecell{Changed\\parameters} \\
    \midrule
    One reader & $0.53$        & \textbf{$-$0.0001} & $0.81$        &  \textbf{511} \\
    Two readers                                & $0.37$        & $-0.0005$          & $0.81$          & $1\,022$ \\
    Three readers (selected)                            & \textbf{0.11} & $-0.0006$          & $0.78$          & $1\,533$ \\
    Three readers, half strength               & $0.30$        & $-0.0004$          & \textbf{0.87} &  $1\,533$ \\
    \bottomrule
  \end{tabular}
  \caption{SAE-guided edits on the smallest model. Columns as in Tab.~\ref{tab:comparison}. The best value of each column is in bold.}
  \label{tab:sae_variants}
\end{table}

Editing more readers removes more of the flaw, from a remaining response of $0.53$ with one reader to $0.11$ with three, while the edits are very small and leave the accuracy untouched. Their weak point is the response to physical changes, which falls to about $0.8$. Removing only half of the direction improves it to $0.87$, but then repairs less.

\paragraph{Edits impact} In Fig.~\ref{fig:patching} (right), two heads besides the three readers, \texttt{b5/5} and \texttt{b6/5}, each carry around $10\%$ of the response, although they give soft neutral particles only $2\%$ of their attention. Activation patching only measures how much of the response passes through a head, not where the particle is first read. The heads of one block run in parallel, and successive blocks all write to the same running representation of each token, so a head can pass on what another head has written. The SAE edits show that this is the case here. \texttt{b5/5} passes on the information that \texttt{b4/0} writes into the particle tokens: editing \texttt{b4/0} reduces its contribution by about three quarters. Similarly, what \texttt{b6/5} writes is no longer used once \texttt{b7/6} and \texttt{c1/2} are edited.

\paragraph{Robustness checks}
We repeat the training of the \texttt{b7/6} SAE with a second random seed and with dictionaries of $512$ to $8\,192$ latents. For dictionaries of $2\,048$ latents or more, both seeds recover the latent, with decoder directions that agree to a cosine of at least $0.93$. With the two smallest dictionaries, one of the two seeds does not recover it. The edit built from the second seed leaves the same remaining response as the first ($0.54$ against $0.53$ for one reader).
\\
Without an SAE, single residual-stream coordinates and single MLP neurons also separate sub-threshold neutral particles from real JetClass neutral particles well on average. They are less clean than the SAE latents when no false positive is allowed: at a threshold that excludes every real JetClass particle, the SAE latents at the three readers still fire on $73\%$, $89\%$ and $82\%$ of the sub-threshold neutral particles, against at most $62\%$ for the single neurons we tested.
\\
For the medium and large models, we train SAEs with the same recipe and find the same latent in both models. We edit three readers in the medium model (\texttt{b7/0}, \texttt{b11/0}, \texttt{b11/7}) and four in the large one (\texttt{b19/11}, \texttt{b18/17}, and the classifier heads \texttt{c0/2} and \texttt{c3/24}). We do not edit \texttt{b11/1} in the medium model, which carries part of the response without attending to the soft particle. Because the response is spread over more heads, the edits leave a remaining response of $0.35$ and $0.58$ (Tab.~\ref{tab:comparison_sizes}).

\subsection{Weight-space programming}
\label{app:wsp}

\paragraph{Agent setup} Each run starts from a folder that contains the model code, the pretrained weights, a description of the jet format, and $10\,240$ practice jets, $1\,024$ per JetClass class. The practice jets come with modified copies: one added photon at $f=10^{-3}$ and at $f=10^{-4}$, and the three physical changes of Eq.~\eqref{eq:ft_phys}. The task description states what may be changed, how an edit is scored, and the rules below. It is the only part that differs between the three variants (see Appendix~\ref{app:control}).

\paragraph{Agent permissions} The agent may change any weight of the body and of the classifier head. It may not change the generator head. It may not compute gradients or use any iterative optimization, and it may not run the model or any of its modules on data. It may use any closed-form computation on the weights and on the activations returned by the evaluation tool, such as singular value decompositions, projections or least-squares fits.

\paragraph{The evaluation tool} All forward passes go through an evaluation tool running on a GPU, which the agent can query at most $50$ times. A query returns one of three things:
\begin{itemize}
    \item the outputs of the original or an edited model on up to $2\,048$ jets, which may be jets the agent has built itself;
    \item the internal activations at up to four places in the network for these jets;
    \item or the scores of a proposed edit on the practice jets.
\end{itemize}

The scores are the response to the added soft particles, the accuracy and the response to physical changes, and the number of changed parameters. The target informed agent can also query the target predictions on any jet, and the scores of its edits include the remaining error. The flaw informed agent is never shown target predictions or its remaining error. With each query, the agent has to state what it expects to see. Before its first query, it has to hand in an edit built from the code and the weights alone.

Each run ends with up to three edits of different sizes and a note that describes the edit, the mechanism the agent believes it has found, the ideas that failed, and its predictions against the outcomes.
\begin{table}[h]
  \centering
  \small
  \setlength{\tabcolsep}{4pt}
  \newcommand{\cell}[2]{\makecell{#1\\{\scriptsize[#2]}}}
  \begin{adjustbox}{max width=\textwidth}
  \begin{tabular}{@{}ll ccc@{}}
    \toprule
    Model & Variant & \makecell{Remaining\\error} & \makecell{Remaining\\response} & \makecell{Changed\\param.} \\
    \midrule
    Small  & Flaw informed           & \cell{$0.33$}{0.26, 0.37} & \cell{$0.14$}{0.10, 0.21} & \cell{$539$}{102, 3\,976} \\
           & Target informed         & \cell{$0.12$}{0.08, 0.23} & \cell{$0.05$}{0.02, 0.06} & \cell{$10\,122$}{142, 10\,267} \\
           & Physics origin informed & \cell{$0.08$}{0.06, 0.09} & \cell{$0.02$}{0.02, 0.04} & \cell{$13\,456$}{491, 14\,321} \\
    \midrule
    Medium & Flaw informed           & \cell{$0.23$}{0.21, 0.33} & \cell{$0.06$}{0.02, 0.07} & \cell{$15\,356$}{1\,185, 29\,701} \\
           & Target informed         & \cell{$0.04$}{0.03, 0.08} & \cell{$0.03$}{0.02, 0.04} & \cell{$17\,564$}{307, 73\,359} \\
           & Physics origin informed & \cell{$0.05$}{0.05, 0.07} & \cell{$0.04$}{0.03, 0.04} & \cell{$5\,628$}{806, 8\,460} \\
    \midrule
    Large  & Flaw informed           & \cell{$0.32$}{0.19, 0.57} & \cell{$0.18$}{0.11, 0.64} & \cell{$14\,849$}{3\,713, 124\,889} \\
           & Target informed         & \cell{$0.10$}{0.07, 0.15} & \cell{$0.04$}{0.04, 0.05} & \cell{$41\,994$}{35\,381, 406\,638} \\
           & Physics origin informed & \cell{$0.19$}{0.07, 0.19} & \cell{$0.15$}{0.04, 0.20} & \cell{$122\,855$}{2\,156, 2\,320\,341} \\
    \bottomrule
  \end{tabular}
  \end{adjustbox}
  \caption{Weight-space programming on the three model sizes. Each cell gives the median over three runs and, below it, the range.}
  \label{tab:wsp_sizes}
\end{table}
\paragraph{Scoring and selection} We score every edit on the test jets of Appendix~\ref{app:repair-method}, which the agent never sees. The task description tells the agent that an edit fails if it lowers the ten-class accuracy by more than $0.010$, the accuracy of any class by more than $0.020$, raises the mean negative log-likelihood by more than $0.030$, or changes the response to one of the physical changes by more than a factor $e^{\pm0.15}$. Each run is represented by the edit with the lowest remaining error among those that pass these limits. The task description also names a limit on the generator head for the small and medium models. We do not use it for the selection, and test the generator separately in Sec.~\ref{sec:heads}.

\paragraph{Results for the three model sizes} Tab.~\ref{tab:wsp_sizes} gives the remaining error, the remaining response and the number of changed parameters for the three variants and the three model sizes.

\paragraph{Cost.} A run uses between $23$ and $50$ of its $50$ queries. It takes between one and two and a half hours for the small model, and up to six hours for the large one, most of it waiting for the evaluation tool. The language-model cost of a run is between $9$ and $16$ US dollars.

\section{Details on the prompts and agent control}
\label{app:control}

\paragraph{Language model} All runs use Claude Opus 5.5~\cite{anthropic2026claudeopus55} through the Claude Code agent, version 2.1.316, between 13 and 17 November 2026. Each run is a separate process started in its own folder, without any memory of earlier sessions and without project instruction files. Before each batch we check, on a run that does nothing else, what the agent sees at start.

\paragraph{Prompts.} Each variant has one task description, which is the same for the three model sizes apart from the facts about the model. The three descriptions share the rules, the scoring and the description of the evaluation tool, and differ in the background and the goal:
\begin{itemize}
    \item \emph{Flaw informed}: the model changes its classification when one very soft neutral particle is added, and the edited model should give the output of the original model on the jet without its very soft neutral particles.
    \item \emph{Target informed}: a reference output exists for every jet, it differs from the original model on some inputs, and the edited model should reproduce it.
    \item \emph{Physics origin informed}: the flaw, its origin in the reconstruction thresholds of JetClass and JetClass-II, the removal rule with its thresholds, and its code.
\end{itemize}
Each description was reviewed, approved and fixed before the first run of its variant, and was not changed afterwards. We identify it by a checksum recorded before launch. The full texts are given in this \href{https://github.com/rbonnetguerrini/WSP-mech-interp-for-model-editing-in-particle-physics}{GitHub repository}.

\paragraph{Isolation} Each run can read and write only its own folder. While a batch of runs is active, the folders of all other runs and of our analysis are made unreadable. The Python interpreter given to the agent records every file it opens, and the agent's commands are recorded in full. After each run, and before looking at any score, we search these records for any access outside the run's folder. We declared before the runs that a run with such an access would be discarded and repeated.

\paragraph{What the records show} No run opened a file of another run or of our analysis, and no run was discarded. We report two events.
\begin{itemize}
    \item In two runs of the flaw informed variant, on the medium and large models, the agent listed the processes of the machine and saw command lines of other agents running at the same time. Both stated that they would ignore them. The ideas concerned also appear in runs that saw nothing. We keep both runs.
    \item Several agents used the system Python interpreter, which is not recorded, for operations on their own files. Their commands remain recorded.
\end{itemize}

\paragraph{Material.} The \href{https://github.com/rbonnetguerrini/WSP-mech-interp-for-model-editing-in-particle-physics}{GitHub repository} contains the three task descriptions with their checksums, the messages used to start the runs, the code of the evaluation tool and of the scoring, all the edits handed in with their scores, the notes and the full transcripts of the agents, and the access records.

\section{Additional results}
\label{app:sizes}
\subsection{Details on the other sizes}
\label{app:sizes_models}
\begin{figure}[h]
  \centering
  \begin{minipage}{0.49\textwidth}\centering
    \includegraphics[width=\textwidth]{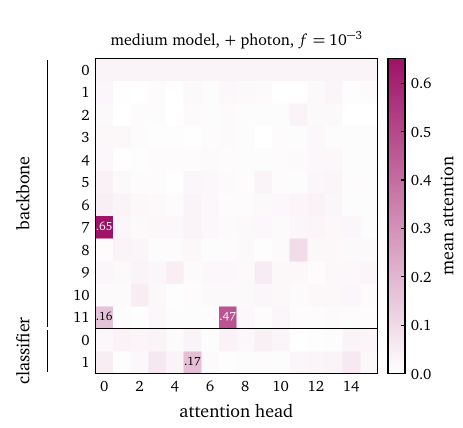}
  \end{minipage}\hfill
  \begin{minipage}{0.49\textwidth}\centering
    \includegraphics[width=\textwidth]{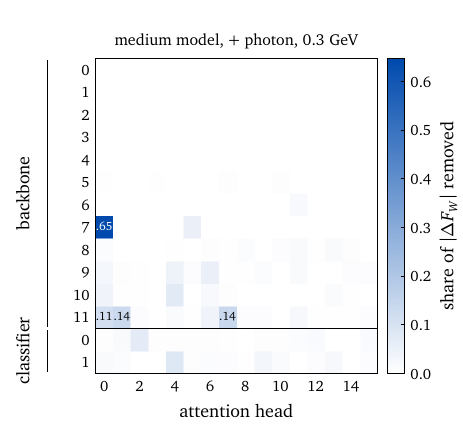}
  \end{minipage}
  \begin{minipage}{0.49\textwidth}\centering
    \includegraphics[width=\textwidth]{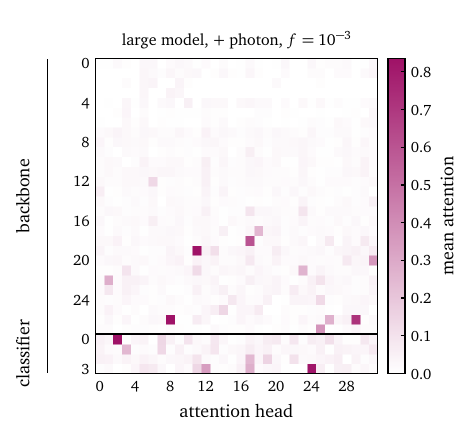}
  \end{minipage}\hfill
  \begin{minipage}{0.49\textwidth}\centering
    \includegraphics[width=\textwidth]{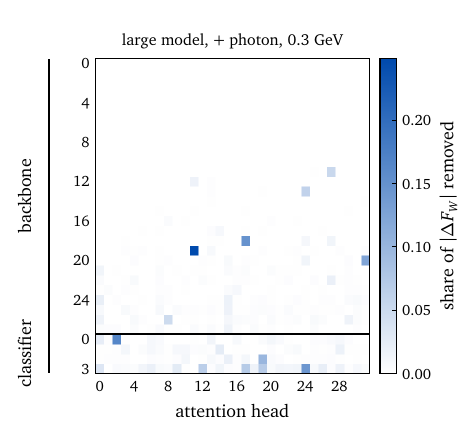}
  \end{minipage}
  \caption{Readers of the medium model (top) and of the large model (bottom). Left: mean attention of the summary tokens on one added photon at $f=10^{-3}$, for every head of the backbone and of the classifier. Right: share of $|\Delta F_W|$ removed when the output of one head is restored to its value on the clean jet, for one added photon at $0.3~\gev$. The color scales differ between the panels.}
  \label{fig:app_readers}
\end{figure}
\paragraph{The flaw} The sensitivity to soft neutral particles is present in the three model sizes, and it is stronger in the two larger ones. After one added photon at $f=10^{-3}$, the $W$ efficiency falls from $80\%$ to $54\%$ for the small model, $28\%$ for the medium model and $40\%$ for the large one. Over the nine taggers, the efficiency falls to between $5\%$ and $29\%$ for the medium model, and between $5\%$ and $42\%$ for the large one.

\paragraph{The readers} Fig.~\ref{fig:app_readers} repeats the analysis of Fig.~\ref{fig:patching} for the two larger models. The medium model resembles the small one: one head, \texttt{b7/0}, carries $65\%$ of the response to a photon at $0.3~\gev$, and three heads of block 11 carry between $11\%$ and $14\%$ each. In the large model no head dominates. The strongest one, \texttt{b19/11}, carries $25\%$, and five others in the last blocks of the body and in the classifier carry between $10\%$ and $17\%$ each.

\paragraph{The repairs} Tab.~\ref{tab:comparison_sizes} gives the comparison of Sec.~\ref{sec:comparison} for the two larger models. Fine-tuning and WSP repair both. The SAE edit loses its effect as the response spreads over more heads: with the readers we identify, it leaves a remaining response of $0.35$ for the medium model and $0.58$ for the large one. WSP changes fewer parameters than fine-tuning on both models in the median, but its edits for the large model range from $35\,000$ to $400\,000$ parameters (Tab.~\ref{tab:wsp_sizes}).

\begin{table}[h]
  \centering
  \small
  \setlength{\tabcolsep}{4pt}
  \begin{adjustbox}{max width=\textwidth}
  \begin{tabular}{@{}ll ccccc@{}}
    \toprule
    Model & Method & \makecell{Remaining\\response} & \makecell{W eff.\\(+1 $\gamma$)} & \makecell{$\Delta$\,acc.\\(clean)} & \makecell{Phys.\\resp.} & \makecell{Changed\\param.} \\
    \midrule
    Medium & Original model & $1$    & $28\%$ & --        & --     & -- \\
           & Fine-tuning    & $0.03$ & $80\%$ & $-0.0005$ & $0.95$ & $354\,640$ \\
           & SAE edit       & $0.35$ & $77\%$ & $+0.0005$ & $0.95$ & $4\,094$ \\
           & WSP            & $0.03$ & $78\%$ & $0.0000$  & $1.00$ & $17\,564$ \\
    \midrule
    Large  & Original model & $1$    & $40\%$ & --        & --     & -- \\
           & Fine-tuning    & $0.05$ & $80\%$ & $0.0000$  & $0.98$ & $1\,593\,568$ \\
           & SAE edit       & $0.58$ & $70\%$ & $+0.0002$ & $0.94$ & $16\,380$ \\
           & WSP            & $0.04$ & $79\%$ & $-0.0001$ & $0.99$ & $41\,994$ \\
    \bottomrule
  \end{tabular}
  \end{adjustbox}
  \caption{Comparison of the repair methods on the medium and large models. Columns as in Tab.~\ref{tab:comparison}. For WSP, the values are the median over the three runs of the selected variant; the ranges are given in Tab.~\ref{tab:wsp_sizes}.}
  \label{tab:comparison_sizes}
\end{table}

\begin{figure}[h]
  \centering
  \includegraphics[width=\textwidth]{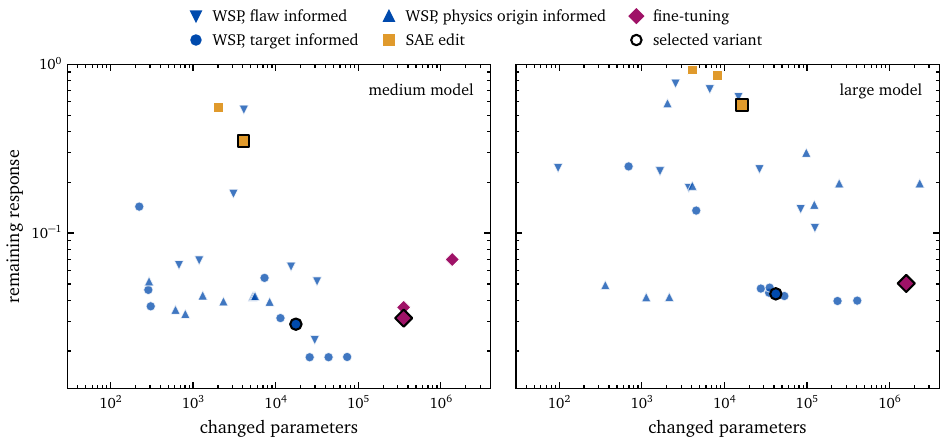}
  \caption{Remaining response against the number of changed parameters, for all the repair variants tested on the medium model (left) and on the large model (right). Each WSP point is one edit of one run, with three edits per run and three runs per variant; Tab.~\ref{tab:wsp_sizes} gives the median and range over the runs. The black outline marks the variant selected for each method (Tab.~\ref{tab:comparison_sizes}); for WSP it marks the median run of the selected variant.}
  \label{fig:app_frontier}
\end{figure}

Fig.~\ref{fig:app_frontier} shows the repair against the size of the change for all the variants tested on the two larger models.

\subsection{The three approaches for all methods and sizes}
\label{app:sizes_tests}

Tab.~\ref{tab:tests_sizes} gives, for each method and model size, the four tests of Sec.~\ref{sec:repaired} after one soft photon is added: the range of the signal efficiency over the nine taggers, the probability that the model assigns to the JetClass-II classes, the change of the prediction of a new mass regression head, and the change of the update predicted by the generator head.
\begin{table}[h]
  \centering
  \small
  \setlength{\tabcolsep}{4pt}
  \newcommand{\cell}[2]{\makecell{#1\\{\scriptsize[#2]}}}
  \begin{adjustbox}{max width=\textwidth}
  \begin{tabular}{@{}ll cccc@{}}
    \toprule
    Model & Method & \makecell{Nine taggers\\efficiency} & \makecell{JetClass-II\\probability} & \makecell{Regression\\head [GeV]} & \makecell{Generator\\update} \\
    \midrule
    Small  & Original model & $12$--$61\%$ & $0.85$ & $10.6$ & $3.3\%$ \\
           & Fine-tuning    & $74$--$79\%$ & $0.06$ & \cell{$5.0$}{4.7, 5.6} & \cell{$1.4\%$}{1.2, 1.5} \\
           & SAE edit       & $75$--$80\%$ & $0.02$ & $1.1$ & \cell{$2.0\%$}{2.0, 2.6} \\
           & WSP            & $78$--$80\%$ & $0.04$ & \cell{$1.1$}{0.5, 1.3} & \cell{$1.0\%$}{1.0, 2.2} \\
           & \emph{Control} & --           & --     & $2.0$  & $1.8\%$ \\
    \midrule
    Medium & Original model & $5$--$29\%$  & $0.88$ & $8.1$  & $3.7\%$ \\
           & Fine-tuning    & $79$--$80\%$ & $0.01$ & $3.0$  & $1.1\%$ \\
           & SAE edit       & $74$--$80\%$ & $0.03$ & $1.5$  & $2.6\%$ \\
           & WSP            & $77$--$79\%$ & $0.04$ & \cell{$0.9$}{0.4, 1.0} & \cell{$3.7\%$}{1.5, 3.7} \\
           & \emph{Control} & --           & --     & $2.9$  & $1.6\%$ \\
    \midrule
    Large  & Original model & $5$--$42\%$  & $0.87$ & $16.4$ & $5.9\%$ \\
           & Fine-tuning    & $79$--$80\%$ & $0.01$ & $1.8$  & $1.9\%$ \\
           & SAE edit       & $15$--$72\%$ & $0.68$ & $6.5$  & $5.0\%$ \\
           & WSP            & $77$--$79\%$ & $0.03$ & \cell{$0.4$}{0.4, 0.4} & \cell{$0.6\%$}{0.2, 3.9} \\
           & \emph{Control} & --           & --     & $1.9$  & $1.9\%$ \\
    \bottomrule
  \end{tabular}
  \end{adjustbox}
  \caption{Tests of Sec.~\ref{sec:repaired} for the three methods and the three model sizes. The first three columns use one added photon at $f=10^{-3}$, the last one a photon of $0.3~\gev$ added to a lightly noised $W$ jet. \emph{Control} is the original model with a photon of $3~\gev$, above the JetClass threshold. Cells with a range give the median and the range over three seeds (fine-tuning, SAE edit) or three runs (WSP).}
  \label{tab:tests_sizes}
\end{table}

The taggers and the regression head are repaired by fine-tuning and WSP on the three model sizes, and by the SAE edit on the small and medium models. On the large model the SAE edit leaves most of the flaw in place, in line with its remaining response.

The generator head is the test on which the methods differ most. Fine-tuning brings its response to the level of the control on the three sizes. For WSP it depends on the run: two of the three runs of the selected variant keep the full response on the medium model, and one of three on the large model. An edit that repairs the classifier outputs does therefore not always repair the representation used by the generator.

\subsection{The response to softer particles}
\label{app:softer}

\begin{figure}[tbp]
  \centering
  \begin{minipage}{0.49\textwidth}\centering
    \includegraphics[width=\textwidth]{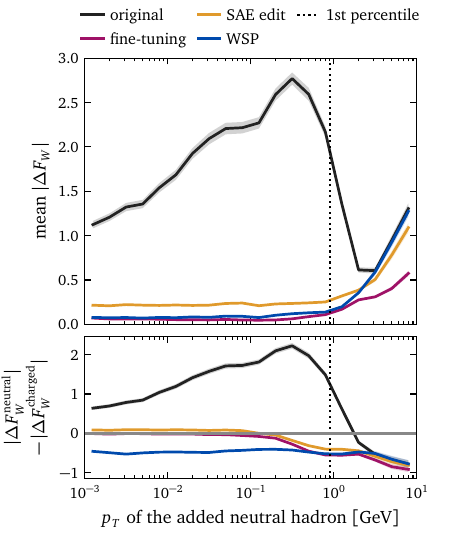}
  \end{minipage}\hfill
  \begin{minipage}{0.49\textwidth}\centering
    \includegraphics[width=\textwidth]{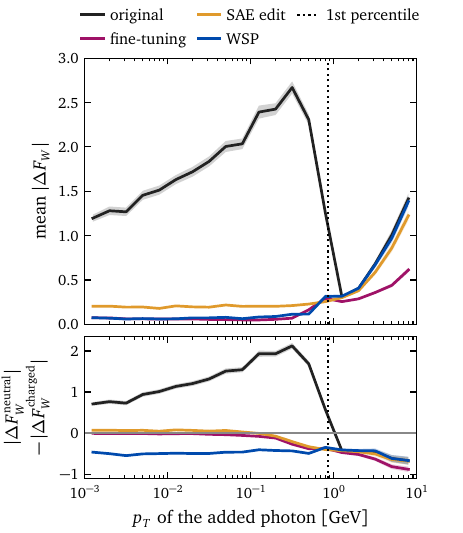}
  \end{minipage}
  \caption{Mean $|\Delta F_W|$ of $W$ jets for one added neutral hadron (left) and photon (right) as a function of its $\pT$, for the original model and the models repaired by fine-tuning, the SAE edit and the selected WSP edit. The lower panels show the difference to the response to one added charged hadron of the same $\pT$, in the same model. The dotted line marks the first percentile of the $\pT$ of the particle type in JetClass jets. $5000$ $W$ jets with two draws each; the bands show the standard error of the mean.}
  \label{fig:ir}
\end{figure}

We test whether the repairs hold for particles much softer than those they were built with, and compare the response to a soft neutral particle with that to a soft charged hadron, to which the original model has no threshold. Fig.~\ref{fig:ir} shows the mean $|\Delta F_W|$ of $W$ jets for one added photon or neutral hadron between $1~\mev$ and $10~\gev$. In the original model the response to a photon rises from $0.3$ around $1~\gev$ to $2.7$ around $0.3~\gev$, below the threshold, and then falls only slowly. At $1~\mev$ it is still $1.2$, more than twice the response to a charged hadron of the same $\pT$. In the three repaired models the response below the threshold is flat, between $0.04$ and $0.23$, down to $1~\mev$. The SAE edit keeps a somewhat larger response to soft neutral particles than fine-tuning and the WSP edit. Fine-tuning and the SAE edit also lower the response to soft charged hadrons, and respond to soft neutral and charged particles alike. The WSP edit leaves the response of the original model to charged hadrons unchanged, and responds less to soft neutral particles than to charged ones. Above the threshold the original model, the WSP edit and the SAE edit respond alike, rising with $\pT$ as expected for a particle that changes the jet. Fine-tuning lowers the response to particles above the threshold as well ($0.55$ instead of $1.4$ for a photon at $8~\gev$).

\subsection{The generator head}
\label{app:generator}

We also measure the relative change of the update the generator predicts for the particles already in the jet, at $t=0.05$, when one photon is added. In the original model a photon at $0.3~\gev$ changes this update by $3.3\%$, against $1.8\%$ for the control photon at $3~\gev$. The repairs lower the response to the sub-threshold photon to about the level of the control: $1.0\%$ for WSP, $1.4\%$ for fine-tuning and $2.0\%$ for the SAE edit (medians over three runs or seeds, with ranges $[1.0, 2.2]$, $[1.2, 1.5]$ and $[2.0, 2.6]$). The change of the generated jet mass, given in the main text, has the ranges $[0.2, 0.4]$ for WSP, $[0.2, 0.3]$ for fine-tuning and $[0.3, 0.6]$ for the SAE edit. A repair of the classifier does not guarantee a repair of the generator though. We observed one run of the physics origin informed variant where the update changes by $3.4\%$ and the mass by $1.3\%$, as much as the original model.

\end{document}